\documentclass[12pt,letterpaper]{article}
\usepackage{amsfonts}
\usepackage{amssymb}
\usepackage{amsmath}
\usepackage{amsthm}
\usepackage{multirow}
\usepackage[round,authoryear]{natbib}
\usepackage[margin=1in]{geometry}
\usepackage{booktabs}
\usepackage{dcolumn}
\newcolumntype{d}[1]{D{.}{.}{#1}} 
\usepackage{graphicx,epstopdf}
\usepackage[T1]{fontenc}
\usepackage{lmodern}
\usepackage{enumitem}
\setlist[enumerate,1]{itemsep=1pt, topsep=4pt, partopsep=0pt}
\setlist[enumerate,2]{nosep}
\setlist[itemize,1]{itemsep=1pt, topsep=4pt, partopsep=0pt}
\setlist[itemize,2]{nosep}
\usepackage{subfig}
\usepackage{bm}
\usepackage{setspace}
\usepackage[colorlinks,linkcolor=red,citecolor=blue,pdftex,breaklinks]{hyperref}
\usepackage[nameinlink]{cleveref}
\hypersetup{bookmarksopen=true}
\theoremstyle{plain}

\theoremstyle{definition}
\Crefname{assumptionx}{Assumption}{Assumptions} %teach cleveref the plural of assumption environment
\Crefname{examplex}{Example}{Examples} %teach cleveref the plural of example environment
\Crefname{remarkx}{Remark}{Remarks} %teach cleveref the plural of remark environment
\makeatletter
\renewcommand*{\eqref}[1]{\hyperref[{#1}]{\textup{\tagform@{\ref*{#1}}}}}
\makeatother

\makeatletter
\DeclareRobustCommand\citepos
  {\begingroup\def\NAT@nmfmt##1{{\NAT@up##1's}}%
   \NAT@swafalse\let\NAT@ctype\z@\NAT@partrue
   \@ifstar{\NAT@fulltrue\NAT@citetp}{\NAT@fullfalse\NAT@citetp}}
\makeatother

\expandafter
\def \expandafter \normalsize \expandafter{\normalsize \setlength \abovedisplayskip{10pt plus 2pt minus 7pt}}
\expandafter
\def \expandafter \normalsize \expandafter{\normalsize \setlength \abovedisplayshortskip{0pt plus 2pt}}
\expandafter
\def \expandafter \normalsize \expandafter{\normalsize \setlength \belowdisplayskip{10pt plus 2pt minus 7pt}}
\expandafter
\def \expandafter \normalsize \expandafter{\normalsize \setlength \belowdisplayshortskip{5pt plus 2pt minus 3pt}}

\def\bbeta{{\bm\beta}}

\def\biX{{\bm{X}}}

\def\biM{{\bm{M}}}

\def\bis{{\bm{s}}}
\def\biy{{\bm{y}}}
\def\bix{{\bm{x}}}
\def\biz{{\bm{z}}}
\def\biu{{\bm{u}}}

\def\bSigma{{\bm{\Sigma}}}

\def\biV{{\bm{V}}}

\def\bdelta{{\bm{\delta}}}

\def\E{{\rm E}}

\def\IF{{\mathbb I}}

\def\N{{\rm N}}

\def\tk{\kern 0.08333em}
\def\tn{\kern -0.08333em}
\def\tkk{\kern 0.04167em}
\def\bzero{{\bm{0}}}
\def\bfI{{\bf I}}

\def\th#1{$#1^{\tk{\rm th}}$}

\def\asysim{\buildrel{a}\over{\sim }}
\def\Mx{\biM_{\tn\biX}}
\def\phm{{\phantom{-}}}

\DeclareMathOperator{\var}{Var}

\begin{document}

\title{When Can We Trust Cluster-Robust Inference?\thanks{A
preliminary version of this paper was presented at the 2025 Canadian
Stata Conference in Ottawa. I received support from the Social
Sciences and Humanities Research Council of Canada (SSHRC grants
435-2021-0396 and 435-2025-1150) and from the Aarhus Center for
Econometrics (ACE) funded by the Danish National Research Foundation
grant number DNRF186. Morten Nielsen, Matt Webb, and two referees
provided useful comments but are not responsible for errors.}}

\author{James G. MacKinnon\thanks{Address: 
Department of Economics, 94 University Avenue, Queen's University, 
Kingston, Ontario K7L 3N6, Canada. Email:\ \texttt{mackinno@queensu.ca}.
Tel.\ 613-533-2293. Fax 613-533-6668.}\\Queen's University\\
Aarhus Center for Econometrics (ACE)\\
\texttt{mackinno@queensu.ca}}

\maketitle

\begin{abstract}

It is common when using cross\tkk-section or panel data to assign each
observation to a cluster and allow for arbitrary patterns of
heteroskedasticity and correlation within clusters. For regression
models, there are many ways to make cluster-robust inferences. A
number of different variance matrix estimators can be used. Hypothesis
tests and confidence intervals can then be based on several
alternative analytic or bootstrap distributions. Some methods
typically perform much better than others, but no method yields
reliable inferences in every case. Thus it can be hard to know which
$P$ values and confidence intervals to trust. Nevertheless, by using a
number of procedures to assess the reliability of various inferential
methods for a specific model and dataset, we can often obtain results
in which we may be reasonably confident.

\end{abstract}

\vskip 12pt



\clearpage

%%%%%%%%%%%%%%%%%%%%%%
\section{Introduction}
\label{sec:intro}
%%%%%%%%%%%%%%%%%%%%%%

In many areas of economics and numerous other disciplines, it is
standard practice to employ cluster-robust standard errors. For
example, between 2021 and 2025, roughly half of the papers in the
\textit{American Economic Review} mentioned the term ``clustered
standard error.'' The proportion was even higher for the
\textit{Quarterly Journal of Economics}.\footnote{These results were
obtained by using Paul Goldman-Pinkham's Economics Literature Search
website \url{https://paulgp.com/econlit-pipeline/search.html}.} It is
natural to use clustered standard errors whenever the observations
fall into disjoint clusters and independence across but not within
clusters can reasonably be assumed. Depending on the application,
clusters might correspond to countries, states, cities, counties,
villages, firms, industries, households, schools or school districts,
hospitals, or time periods. This list is, of course, by no means
exhaustive.

The idea of cluster-robust standard errors is a natural generalization
of the concept of heteroskedasticity-consistent ones
\citep{White_1980}. Indeed, several methods for calculating the former
reduce to methods for calculating the latter when all clusters contain
just one observation; see \Cref{sec:linreg}. Cluster-robust standard
errors were first studied, for the special case of equal-sized
clusters, in \citet{White_1984}. Subsequent theoretical work includes
\citet{LZ_1986}, \citet{Arellano_1987}, \citet*{BCH_2011},
\citet{HansenLee_2019}, \citet*{DMN_2019}, and \citet{Hansen-jack}.
All of these works provide asymptotic results under various
assumptions. In particular, the two 2019 papers obtain asymptotic
distributions for cluster-robust $t$-statistics while allowing both
the number of clusters and the number of observations per cluster to
increase with the sample size.

Although asymptotic theory is essential and can be illuminating, it
usually only tells us that cluster-robust inference can be relied upon
when the number of clusters goes to infinity. Unfortunately, there are
many circumstances in which inference can be seriously unreliable when
the number of clusters is not large. These will be discussed in
\Cref{sec:unreliable}. Although much is known about what causes
various inferential methods to perform well or badly in finite samples
\citep*{MNW-guide}, what we know does not generally allow us to be
sure that any particular method can, or cannot, be trusted for any
particular model and dataset. This is a rather unsatisfactory state of
affairs.

Nevertheless, if one is willing to compute several different test
statistics and $P$~values, along with some auxiliary diagnostics, it
is often possible to get a good idea of which results to trust. The
main contribution of the paper is to discuss a number of practical
procedures and illustrate their use. \Cref{sec:linreg} reviews the
linear regression model, cluster-robust variance matrices, and several
analytical methods of inference. \Cref{sec:bootstrap} then discusses
bootstrap inference. \Cref{sec:unreliable} briefly explains when and
why inference may be unreliable, and \Cref{sec:how} discusses how to
determine the correct way in which to cluster. \Cref{sec:assess}
discusses a number of procedures for assessing the reliability of
various inferential methods, including targeted Monte Carlo
experiments and placebo regressions. \Cref{sec:applic} illustrates
these procedures in the context of two empirical applications, and
\Cref{sec:conclusion} concludes.

%%%%%%%%%%%%%%%%%%%%%%%%%%%%%%%%%%%%%%%%%%%
\section{Linear Regression with Clustering}
\label{sec:linreg}
%%%%%%%%%%%%%%%%%%%%%%%%%%%%%%%%%%%%%%%%%%%

Although many types of statistical model can have clustered
disturbances, I focus on the linear regression model with one\tkk-way
clustering. This model is conceptually and computationally easy to
deal with, and it is very commonly encountered in practice. Suppose
there are $G$ clusters, indexed by $g$. The \th{g} cluster has $N_g$
observations, and the sample size is $N = \sum_{g=1}^G N_g$. The model
can be written as
\begin{equation}
\label{eq:lrmodel} 
\biy_g =\biX_g\bbeta + \biu_g, \quad g=1,\ldots,G.
\end{equation}
Here $\biX_g$ is an $N_g\times k$ matrix of regressors, $\bbeta$ is a
$k$-vector of coefficients, $\biy_g$ is an $N_g$-vector of
observations on the regressand, and $\biu_g$ is an $N_g$-vector of
disturbances (or error terms). Stacking the $\biy_g$ yields the
$N$-vector $\biy$, stacking the $\biX_g$ yields the $N\times k$ matrix
$\biX$\tn, and stacking the $\biu_g$ yields the $N$-vector $\biu$, so
that \eqref{eq:lrmodel} can be rewritten as $\biy = \biX\tn\bbeta +
\biu$.

The ordinary least squares (OLS) estimator of $\bbeta$ is
\begin{equation}
\hat\bbeta = (\biX^\top\!\biX)^{-1}\biX^\top\biy
= \bbeta_0 +  (\biX^\top\!\biX)^{-1}\biX^\top\biu,
\label{eq:OLSbeta}
\end{equation}
where the second equality in \eqref{eq:OLSbeta} depends on the
assumption that the data are generated by \eqref{eq:lrmodel} with
$\bbeta = \bbeta_0$. Thus, if $\bis_g = \biX_g^\top \biu_g$ is the
score vector for the \th{g} cluster,
\begin{equation}
\hat\bbeta - \bbeta_0 = (\biX^\top\!\biX)^{-1}\sum_{g=1}^G
\biX_g^\top\biu_g
= \Big(\tn\sum_{g=1}^G\biX_g^\top\!\biX_g\Big)^{\!\!-1} \sum_{g=1}^G
\bis_g.
\label{eq:betahat}
\end{equation}
All of the random variation in $\hat\bbeta$ around $\bbeta_0$
evidently comes from the randomness in the score vectors. In order to
make inferences about $\bbeta$, we need to make assumptions about how
these vectors are distributed. It is conventional to assume,
conditional on~$\biX$\tn, that
\begin{equation}
\label{eq:Sigma_g}
\E(\bis_g\bis_g^\top) = \bSigma_g \quad\mbox{and}\quad
\E(\bis_g\bis_{g'}^\top) = \bzero, \quad g,g'=1,\ldots,G,\quad g'\ne g,
\end{equation}
where $\bSigma_g$ is the symmetric, positive semidefinite variance
matrix of the scores for the \th{g} cluster. The first assumption in
\eqref{eq:Sigma_g} allows for very general patterns of
heteroskedasticity and intra-cluster correlation within each cluster.
The second assumption is stronger, and it is crucial. It states that
the scores for every cluster are uncorrelated with the scores for
every other cluster. When the second assumption is not true, all the
methods to be discussed in this paper are asymptotically invalid. Thus
it can be important to test it, unless there is really only one
sensible way to cluster; see \Cref{subsec:level}.

From \eqref{eq:Sigma_g} and the rightmost expression in
\eqref{eq:betahat}, the true variance matrix of $\hat\bbeta$,
conditional on $\biX$\tn, is seen to be the sandwich matrix
\begin{equation}
\label{eq:trueV}
\var(\hat\bbeta) =
(\biX^\top\!\biX)^{-1} \Big(\tn\sum_{g=1}^G \bSigma_g\tn\Big) 
(\biX^\top\!\biX)^{-1}.
\end{equation}
Somewhat non-standard asymptotic arguments \citep*[Theorem
2.1]{DMN_2019} then allow us to proceed as if
\begin{equation}
\label{eq:asybeta}
\hat\bbeta \asysim \N\big(\bbeta_0, \var(\hat\bbeta)\big).
\end{equation}
If we knew $\var(\hat\bbeta)$, it seems plausible that
\eqref{eq:asybeta} would allow us to make reliable inferences, except
perhaps when $G$ is very small, or the score vectors are very
heterogeneous, so that the central limit theorem needed for
\eqref{eq:asybeta} does not perform well.

In practice, of course, we do not know $\var(\hat\bbeta)$, because we
do not know the $\bSigma_g$. We have to use some estimator
$\widehat\var(\hat\bbeta)$. This involves replacing the $\bSigma_g$ in
\eqref{eq:trueV} by consistent estimates $\hat\bSigma_g$. There is
more than one way to do this. Unfortunately, all of them can lead to
seriously unreliable inferences in some cases.

%%%%%%%%%%%%%%%%%%%%%%%%%%%%%%%%%%%%%%%%%%%%%
\subsection{Three Variance Matrix Estimators}
\label{subsec:CRVEs}
%%%%%%%%%%%%%%%%%%%%%%%%%%%%%%%%%%%%%%%%%%%%%

There are several ways to estimate the middle factor in
\eqref{eq:trueV}. The simplest approach is to replace $\bSigma_g$ by
$\hat\bis_g\hat\bis_g^\top$, where $\hat\bis_g = \biX_g^\top
\hat\biu_g$ is the empirical score vector for the \th{g} cluster. If
we also multiply the entire matrix by a correction for degrees of
freedom, we obtain
\begin{equation}
\mbox{CV$_{\tn1}$:}\qquad \hat\biV_1(\hat\bbeta) =
\frac{G(N-1)}{(G-1)(N-k)}\tk
(\biX^\top\!\biX)^{-1}
\Big(\tn\sum_{g=1}^G \hat\bis_g\hat\bis_g^\top\Big)
(\biX^\top\!\biX)^{-1}.
\label{eq:CV1}
\end{equation}
This is by far the most widely-used cluster-robust variance matrix
estimator, or CRVE, in practice. The leading scalar is chosen so that,
when $G=N$\tn, $\hat\biV_1(\hat\bbeta)$ reduces to the familiar HC$_1$
estimator of \citet{MW_1985} that is robust to heteroskedasticity of
unknown form. The estimator in \eqref{eq:CV1} is therefore commonly
referred to as CV$_{\tn1}$.

The empirical score vectors $\hat\bis_g$ do not always provide good
estimates of the actual score vectors~$\bis_g$, because the residual
vectors $\hat\biu_g$ do not always provide good estimates of the
disturbance vectors~$\biu_g$. In general, $\hat\biu = \Mx\biu$, where
$\Mx = \bfI - \biX(\biX^\top\biX)^{-1}\biX^\top$ is an $N\times N$
orthogonal projection matrix. Depending on the properties of $\Mx$
and~$\biu$, the~$\hat\bis_g$ can sometimes differ greatly from the
$\bis_g$, causing the middle factor in \eqref{eq:CV1} to provide a
poor estimate of $\sum_{g=1}^G \bSigma_g$. This implies that the
CV$_{\tn1}$ variance estimator \eqref{eq:CV1} may not always perform
well, and it also suggests alternative estimators that generally
perform better.

Two widely-used alternatives to CV$_{\tn1}$ transform the
$\hat\biu_g$, and hence also the empirical score vectors, before
estimating the middle factor in \eqref{eq:trueV}. These CRVEs are
generally known as CV$_{\tn2}$ and CV$_{\tn3}$, because they are
analogous to the HC$_2$ and HC$_3$ estimators of \citet{MW_1985}. For
HC$_2$, the residuals $\hat u_i$ are replaced by the rescaled
residuals $\hat u_i/M_{ii}^{1/2}$, and for HC$_3$ they are replaced by
$\hat u_i/M_{ii}$, where $M_{ii}$ is the \th{i} diagonal element
of~$\Mx$. The CV$_{\tn2}$ and CV$_{\tn3}$ estimators were originally
proposed in \citet{BM_2002}. In that paper, they were calculated using
algebraic procedures analogous to the ones for HC$_2$ and HC$_3$.
These involve, respectively, the inverse symmetric square roots or the
inverses of the $\biM_{gg}$, which are the diagonal blocks of $\Mx$
corresponding to each cluster. As a result, these procedures are
extremely costly (perhaps computationally infeasible) when any of the
clusters are large. Much better computational procedures for
CV$_{\tn2}$ and CV$_{\tn3}$, some of them due to \citet{NAAMW_2020},
are discussed in \citet*{MNW-bootknife}.

The CV$_{\tn3}$ variance matrix is particularly interesting because,
like HC$_3$, it is based on the jackknife
\citep{Tukey_1958,Efron-Stein}. In fact, it is a form of
cluster-jackknife estimator. The idea of the cluster jackknife is to
compute $G$ sets of parameter estimates, each of which omits one
cluster at a time. The OLS estimates of $\bbeta$ when each cluster is
omitted in turn are
\begin{equation}
\label{eq:delone}
\hat\bbeta^{(g)} = (\biX^\top\!\biX - \biX_g^\top\!\biX_g)^{-1}
(\biX^\top\biy - \biX_g^\top\biy_g), \quad g=1,\ldots,G.
\end{equation}
It is easy to obtain the $\hat\bbeta^{(g)}$ in a computationally
efficient manner. Start by calculating the cluster-level matrices
and vectors
\begin{equation}
\biX_g^\top\!\biX_g \quad\mbox{and}\quad \biX_g^\top\biy_g, \quad
g=1,\ldots,G.
\label{eq:subthings}
\end{equation}
Unless $G$ is very large, this involves very little cost beyond that
of computing $\hat\bbeta$, because the quantities in
\eqref{eq:subthings} can be used to construct $\biX^\top\!\biX$ and
$\biX^\top\biy$. It is evident from \eqref{eq:delone} that the main
cost, after $\hat\bbeta$ and its ingredients have been computed, is
calculating the inverse (or possibly the generalized inverse) of a
$k\times k$ matrix for each of the $\hat\bbeta^{(g)}$\tn. Unless $k$
and/or $G$ are very large, this should not be prohibitively expensive.

The simplest version of the cluster-jackknife variance matrix is
\begin{equation}
\mbox{CV$_{\tn3}$:}\qquad
\hat\biV_3(\hat\bbeta) = \frac{G-1}{G} \sum_{g=1}^G
(\hat\bbeta^{(g)} - \hat\bbeta)(\hat\bbeta^{(g)} - \hat\bbeta)^\top\tn.
\label{eq:CV3}
\end{equation}
This matrix is usually not difficult or expensive to compute, although
it may be necessary to deal with singularities in the matrices
$\biX^\top\!\biX - \biX_g^\top\!\biX_g$. \texttt{Stata} can compute it
using either the \texttt{vce(jackknife,mse)} option or the
\texttt{summclust} package; see \Cref{subsec:hetero}.

\citet{Hansen-jack} contains some interesting theoretical results on
the CV$_{\tn3}$ estimator and modified versions of it. These suggest
that standard errors based on it will almost always be larger than
ones based on CV$_{\tn1}$. Simulation evidence in
\citet*{MNW-bootknife,MNW-influence} strongly suggests that
$t$-statistics based on CV$_{\tn3}$ almost always yield more reliable
inferences than ones based on CV$_{\tn1}$. In cases where reliable
inference is difficult (see \Cref{sec:unreliable}), the gain from
using CV$_{\tn3}$ instead of CV$_{\tn1}$ can be substantial.

%%%%%%%%%%%%%%%%%%%%%%%%%%%%%%%%%%
\subsection{Inference Using CRVEs}
\label{subsec:inference}
%%%%%%%%%%%%%%%%%%%%%%%%%%%%%%%%%%

Suppose we are interested in a single element of the vector~$\bbeta$,
say the coefficient~$\beta_j$. We might wish to test the hypothesis
that $\beta_j=\beta_{0j}$, where $\beta_{0j}$ is a known value, or we
might wish to construct a confidence interval for $\beta_j$. In either
case, it seems natural to start with the $t$-statistic
\begin{equation}
t^m_j = \frac{\hat\beta_j - \beta_{0j}}{\textrm{se}_m(\hat\beta_j)},
\quad m=1,2,3,
\label{eq:cr-tstat}
\end{equation}
where se$_m(\hat\beta_j)$ is the square root of the \th{j} diagonal
element of CV$_{\tn m}$. By itself, of course, a $t$-statistic does
not allow us to make inferences. We need to assume that it follows
some distribution. By combining \eqref{eq:cr-tstat} with an assumed
distribution, we can perform hypothesis tests or construct confidence
intervals. But what distribution should we assume?

Because the asymptotic theory that justifies \eqref{eq:asybeta} and
all three CRVEs merely tells us that $\hat\bbeta$ is asymptotically
normal and that each of the CRVEs validly estimates $\var(\hat\bbeta$)
as $G\to\infty$, it might seem natural to employ the standard normal
distribution. However, there is an alternative asymptotic theory which
yields a different result. In this theory, which is due to
\citet*{BCH_2011}, $G$ is held fixed, and the number of observations
per cluster is allowed to increase with~$N$ while the amount of
intra-cluster correlation is assumed to diminish. For CV$_{\tn1}$,
this theory yields the striking result that the $t$-statistic
\eqref{eq:cr-tstat} is asymptotically distributed as $t(G-1)$.

Since there are G matrices $\bSigma_g$ to be estimated, it seems
plausible that $t^m_j$ should follow the $t(G-1)$ distribution as an
approximation for all three CRVEs. Thus, although there are to my
knowledge no theoretical results like those of \citet*{BCH_2011} to
justify the use of the $t(G-1)$ distribution with CV$_{\tn2}$ and
CV$_{\tn3}$ standard errors, it is conventional to use this
distribution with $t$-statistics based on any of the three CRVEs.

What about CV$_{\tn2}$ standard errors? They have the attractive
feature that, when there is neither heteroskedasticity nor
intra-cluster correlation, and certain assumptions on $\biX$ are
satisfied, the diagonal elements of CV$_{\tn2}$ are unbiased
\citep{BM_2002}, like the diagonal elements of HC$_2$. In contrast,
the diagonal elements of CV$_{\tn3}$ and HC$_3$ are generally biased
upwards in the special case of i.i.d.\ disturbances. Unfortunately,
using the square root of an unbiased variance estimator in the
denominator of a $t$-statistic does not guarantee that it will follow
a $t$ distribution. This would require the numerator to be normally
distributed, the denominator to be the square root of an appropriately
scaled $\chi^2$ random variable, and the numerator to be independent
of the denominator. None of these conditions holds here.

In fact, simulation evidence in \citet*{MNW-bootknife} suggests that
using CV$_{\tn3}$ $t$-statistics in conjunction with the $t(G-1)$
distribution is more reliable than using CV$_{\tn2}$ $t$-statistics,
which in turn is more reliable than using CV$_{\tn1}$ $t$-statistics.
Only in a few cases, where CV$_{\tn3}$ $t$-statistics actually tend to
under-reject (see \Cref{sec:unreliable}), does it seem to be
preferable to use CV$_{\tn2}$. It never seems to be preferable to use
CV$_{\tn1}$.

The $t(G-1)$ distribution is only an approximation. Some papers
suggest using a calculated degrees\tkk-of-freedom parameter, say
$d_j$, in place of $G-1$. These include \citet{BM_2002},
\citet{Imbens_2016}, \citet{AY-exact}, and
\citet{Hansen-jack,Hansen_2025}. There are several different
expressions for $d_j$, which depend on which CRVE is being used and on
what simplifying assumptions are made about the disturbances. The
subscript on $d_j$ emphasizes the fact that the calculated d-o-f
parameter is specific to the coefficient~$\beta_j$.

Except in very special circumstances, all CRVEs are biased. This bias
can be corrected if the \th{j} diagonal element of the CRVE is
multiplied by a suitable scaling factor, say $\gamma_j$. Papers that
take this approach include \citet{AY-exact}, \citet*{BNW_2023}, and
\citet{Hansen-jack,Hansen_2025}. Once again, the scaling factor is
specific to the coefficient~$\beta_j$.

Of the methods mentioned above, the best seem to be ones that use both
a scaling factor and a calculated d-o-f parameter. A good deal of
algebra is needed to calculate the $d_j$ and the $\gamma_j$, and this
needs to be repeated for every coefficient of interest. Fortunately,
the method of \citet{Hansen-jack,Hansen_2025} is now available in
Stata 19.5.

%%%%%%%%%%%%%%%%%%%%%%%%%%%%%
\section{Bootstrap Inference}
\label{sec:bootstrap}
%%%%%%%%%%%%%%%%%%%%%%%%%%%%%

An alternative approach, which often works very well, is to employ
bootstrap methods. The idea is to generate a large number of bootstrap
samples, say $B$, indexed by~$b$. For each of them, we compute a
vector of bootstrap parameter estimates, say $\hat\bbeta^{*b}$\tn. In
most cases, we also compute other quantities, such as bootstrap test
statistics. Ideally, the $\hat\bbeta^{*b}$ should have approximately
the same distribution as $\hat\bbeta$, and the bootstrap test
statistics should have approximately the same distribution(s) as the
actual test statistics to which they correspond. If so, we can use the
bootstrap distributions to make inferences.

%%%%%%%%%%%%%%%%%%%%%%%%%%%%%%%%%%%%%%%%
\subsection{The Pairs Cluster Bootstrap}
\label{subsec:pairs}
%%%%%%%%%%%%%%%%%%%%%%%%%%%%%%%%%%%%%%%%

The best-known, and conceptually the simplest, bootstrap method for
models like \eqref{eq:lrmodel} is the pairs cluster bootstrap, or PCB.
It is a natural extension of the original resampling bootstrap of
\citet{Efron_79}. Instead of resampling individual observations, it
resamples clusters of data on all the variables. For the model
\eqref{eq:lrmodel}, there is no need to form actual bootstrap samples.
Recall that, in order to calculate CV$_{\tn3}$ efficiently, we need to
compute the matrices and vectors $\biX_g^\top\biX_g$ and
$\biX_g^\top\biy_g$ for each cluster. If we resample directly from
pairs of these, the \th{b} bootstrap sample then consists of
\begin{equation}
\label{bpairs}
\left[{\biX_g^{*b}}^\top\!\biX_g^{*b},\;
{\biX_g^{*b}}^\top\!\biy_g^{*b}\right], \quad g=1,\ldots,G,
\end{equation}
where each of the bootstrap pairs in \eqref{bpairs} equals one of the
original pairs $[\biX_g^\top\biX_g,\, \biX_g^\top\biy_g]$ with
probability $1/G$. The bootstrap estimate of $\bbeta$ is then
\begin{equation*}
%\label{pbboot}
\hat\bbeta^{*b} = \Bigl(\sum_{g=1}^G
{\biX_g^{*b}}^\top\!\biX_g^{*b}\Big)^{\!\!-1}
\sum_{g=1}^G {\biX_g^{*b}}^\top\!\biy_g^{*b} =
\big({\biX^{*b}}^\top\!\biX^{*b}\big)^{\!-1}
{\biX^{*b}}^\top\!\biy^{*b}.
\end{equation*}
Since it is straightforward to calculate the variance matrices of
\Cref{subsec:CRVEs} for each of the bootstrap samples using only the
matrices and vectors in \eqref{bpairs} \citep{JGM-fast}, we can
also compute the bootstrap $t$-statistics
\begin{equation*}
t_j^{*b} = \frac{\hat\beta^{*b}_j - \hat\beta_j}
{\textrm{se}(\hat\beta^{*b}_j)},\quad b=1,\ldots,B,
%\label{eq:pb-tstat}
\end{equation*}
for any choice of se$(\hat\beta^{*b}_j)$. Notice that the bootstrap
$t$-statistic here is testing the hypothesis that \smash{$\beta_j =
\hat\beta_j$}, not the hypothesis that $\beta_j = \beta_{0j}$. Because
the bootstrap samples do not impose any restrictions on $\bbeta$, the
latter hypothesis is not true for them.

As usual, the bootstrap quantities $\hat\beta^{*b}_j$ and $t^{*b}_j$
can be used to make inferences about $\beta_j$ in more than one way.
One possibility is to compute the bootstrap standard error of
\smash{$\hat\beta_j$}, say se$^*(\hat\beta_j)$, as the square root of
the sample variance of the $\hat\beta_j^{*b}$. In cases where
inference is easy, se$^*(\hat\beta_j)$ should be similar to all of the
\smash{se$_m(\hat\beta_j)$}. However, both theory and simulations
suggest that, instead of using bootstrap standard errors, it is
usually better to test the hypothesis that $\beta_j = \beta_{0j}$ by
computing the symmetric bootstrap $P$ value
\begin{equation}
\hat{P}^* = \frac{1}{B}\sum_{b=1}^{B}
\IF\bigl(|t^{*b}_j| > |t_j|\bigr),
\label{symbootP}
\end{equation}
where $\IF(\cdot)$ is the indicator function, which is 1 if its
argument is true and 0 otherwise. When $\hat{P}^* < \alpha$, we can
reject the null hypothesis at level~$\alpha$. Upper-tail or equal-tail
$P$ values could also be used, but when $\hat\bbeta$ is not biased,
the symmetric one seems to work slightly better.

It is also straightforward to compute studentized bootstrap confidence
intervals. Sort the $t^{*b}_j$ from smallest to largest and find the
empirical $1-\alpha/2$ and $\alpha/2$ quantiles, say
$c^*_{1-\alpha/2}$ and $c^*_{\alpha/2}$, respectively. For example, if
$B=999$ and $\alpha=0.05$, then $c^*_{\alpha/2}$ would be number 25
and $c^*_{1-\alpha/2}$ would be number 975 in the sorted list of the
$t^{*b}_j$. The studentized bootstrap confidence interval at level
$1-\alpha$ is then simply
\begin{equation}
\big[\hat\beta_j - {\rm se}(\hat\beta_j) c^*_{1-\alpha/2}, \quad
\hat\beta_j - {\rm se}(\hat\beta_j) c^*_{\alpha/2}\big],
\label{eq:studboot}
\end{equation}
where the cluster-robust standard error se$(\hat\beta_j)$ is the same
function of the actual data as se$(\hat\beta^{*b}_j)$ is a function of
the bootstrap data.

Although the PCB is asymptotically valid for a wide set of econometric
models, simulation results suggest that it often does not work
particularly well \citep*{CGM_2008,MW-TPM}. One reason for its
mediocre performance is that each bootstrap sample contains a
different set of resampled clusters. Thus, unless all clusters are the
same size, the bootstrap sample sizes will vary. Moreover, when the
$\biX_g^\top\biX_g$ matrices vary greatly among themselves, so that
some clusters have much higher leverage than others, some of the
\smash{${\biX^{*b}}^\top\!\biX^{*b}$} matrices may not look much like
$\biX^\top\biX$\tn. Thus it may not be safe to pretend that $t_j$ is a
drawing from the distribution of the $t^{*b}_j$.

%%%%%%%%%%%%%%%%%%%%%%%%%%%%%%%%%%%%%%%
\subsection{The Wild Cluster Bootstrap}
\label{subsec:wild}
%%%%%%%%%%%%%%%%%%%%%%%%%%%%%%%%%%%%%%%

A bootstrap method that often works better than the PCB is the wild
cluster bootstrap, or WCB. It was originally proposed in
\citet*{CGM_2008}, and its asymptotic validity was proved in
\citet*{DMN_2019}. Several new variants of the WCB that generally work
better than the original were developed in \citet*{MNW-bootknife}.
The most appealing of these will be discussed below.

Most expositions of the WCB involve residuals and fitted values, but
it is conceptually and computationally easier to formulate it in terms
of the score vectors. Consider the unrestricted empirical score vectors
\begin{equation}
\label{eq:score}
\hat\bis_g = \biX_g^\top\hat\biu_g = \biX_g^\top\biy_g -
\biX_g^\top\biX_g\tk\hat\bbeta, \;\; g=1,\ldots,G.
\end{equation}
To generate the \th{b} bootstrap sample, we multiply each of these
vectors by a scalar auxiliary random variable $v_g^{*b}$ with mean~0
and variance~1. Unless $G$ is very small \citep{Webb_six}, the best
choice for the distribution of $v_g^*$ seems to be the Rademacher
distribution, which takes the values 1 and $-1$ with equal probability
\citep*{DF_2008,DMN_2019}.

We then obtain bootstrap estimates
\begin{equation}
\label{bootbeta}
\hat\bdelta^{*b} = (\biX^\top\biX)^{-1} \sum_{g=1}^G\bis_g^{*b}, \quad
\bis_g^{*b} = v_g^{*b}\hat\bis_g,
\end{equation}
where $\hat\bdelta^{*b} \equiv \hat\bbeta^{*b} - \hat\bbeta$.
Next, we compute the bootstrap $t$-statistic
\begin{equation}
\label{boottk}
t_j^{*b} = \frac{\hat\delta_j^{*b}}{{\rm se}(\hat\delta_j^{*b})}\tk,
\end{equation}
where ${\rm se}(\hat\delta_j^{*b})$ is the square root of the \th{j}
diagonal element of the CV$_{\tn1}$ matrix \eqref{eq:CV1}, with the
$\hat\bis_g$ replaced by \smash{$\hat\bis^{*b}_g = \bis^{*b}_g -
\biX_g^\top\biX_g \hat\bdelta^{*b}$}. The $t_j^{*b}$ can then be used
to compute bootstrap $P$ values, using \eqref{symbootP}, and
studentized bootstrap confidence intervals, using \eqref{eq:studboot},
in exactly the same way as for the PCB. This variant of the WCB is
called the WCU-C bootstrap. The ``U'' indicates that the bootstrap
scores are based on the unrestricted empirical score vectors
\eqref{eq:score}, and the ``-C'' stands for ``classic'' to distinguish
it from newer methods.

Imposing restrictions on the bootstrap samples makes bootstrapping
somewhat more complicated, but it often improves the finite\tkk-sample
properties of bootstrap tests \citep{DM_1999}. Under the restriction
$\beta_j=\beta_{j0}$, OLS yields restricted estimates $\tilde\bbeta$
and restricted residual vectors $\tilde\biu_g$, for $g=1,\ldots,G$.
Then the analog of \eqref{eq:score} is
\begin{equation}
\label{eq:rscore}
\tilde\bis_g = \biX_g^\top\tilde\biu_g = \biX_g^\top\biy_g -
\biX_g^\top\biX_g\tk\tilde\bbeta, \;\; g=1,\ldots,G.
\end{equation}
Because one element of $\tilde\bbeta$ is $\beta_{j0}$, the vectors
$\tilde\bis_g$ have $k$ elements, even though only $k-1$ parameters
were estimated. The formulae for \smash{$\hat\bdelta^{*b}$} and
\smash{$t_j^{*b}$} are once again \eqref{bootbeta} and \eqref{boottk},
but now $\hat\bdelta^{*b} = \hat\bbeta^{*b} - \tilde\bbeta$.

This variant of the WCB is called the WCR-C variant. Here ``R'' stands
for ``restricted,'' because the bootstrap scores are based on the
restricted empirical score vectors \eqref{eq:rscore}. This is the
procedure recommended in \citet*{CGM_2008}, \citet*{DMN_2019}, and
many other papers. Bootstrap $P$ values can once again be computed
using \eqref{symbootP}. However, because the bootstrap samples are
generated subject to a restriction on $\beta_j$, we cannot construct
studentized bootstrap confidence intervals using \eqref{eq:studboot}.
Instead, we have to ``invert'' the bootstrap test statistic, finding
two values of $\beta_j$, one on each side of $\hat\beta_j$, such that
the equal-tail $P$ values for tests that $\beta_j$ equals each of
these values are approximately~$\alpha$. For details, see
\citet[Section~3.4]{JGM-fast}.

Two of the new variants proposed in \citet*{MNW-bootknife} are
particularly easy to compute and seem to perform better than the
classic ones in many cases. These variants are known as WCU-S and
WCR-S, where ``S'' stands for ``score.'' The idea of the ``score''
wild cluster bootstraps is to replace the empirical score vectors
$\hat\bis_g$ or $\tilde\bis_g$ in the bootstrap data-generating
process (DGP) by modified score vectors that at least partly correct
for the distortions caused by least squares. Recall that $\hat\biu =
\Mx\biu$. We cannot recover~$\biu$ by multiplying $\hat\biu$ by the
inverse of $\Mx$, because $\Mx$ is a singular matrix and often a very
large one. However, we can do something along the same lines on a
cluster-by-cluster basis. Assuming that $\biM_{gg}$, the \th{g}
diagonal block of $\Mx$\tn, is invertible, we can define the modified
score vector
\begin{equation}
\acute\bis_g = \biX_g^\top\biM_{gg}^{-1}\tk\hat\biu_g.
\label{eq:bis3}
\end{equation}
Using \eqref{eq:bis3} is expensive, or even computationally
infeasible, for large clusters. However, it is shown in
\citet*{MNW-bootknife} that
\begin{equation*}
\acute\bis_g = \biX^\top\biX\big(\hat\bbeta - \hat\bbeta^{(g)}\big),
\;\; g=1,\ldots,G.
\end{equation*}
Thus computing the $\acute\bis_g$ is almost costless once the
jackknife estimates, which are needed for CV$_{\tn3}$, have been
computed. The modified score vectors $\acute\bis_g$ are used in the
bootstrap DGP for the WCU-S bootstrap. In all other respects, the
WCU-S and WCU-C bootstraps are computed in exactly the same way.

A similar, but somewhat more complicated, procedure can be used to
compute restricted score vectors $\dot\bis_g$ that ``correct'' for the
distortions caused by estimating the restricted model; see
\citet*[Section~5]{MNW-bootknife}. The $\dot\bis_g$ are used in the
bootstrap DGP for the WCR-S bootstrap, which otherwise is computed in
exactly the same way as the WCR-C bootstrap.

It may seem odd to use the CV$_{\tn1}$ standard error in $t_j$ and the
$t_j^{*b}$\tn, when the transformation \eqref{eq:bis3} is based on the
cluster jackknife. There is another variant of the wild cluster
bootstrap that uses the CV$_{\tn3}$ standard error, but it is a lot
more expensive to compute, and it does not consistently perform better
\citep*[Section~6]{MNW-bootknife}.

The four wild bootstrap methods discussed in this section are
implemented in current versions of the \texttt{Stata} package
\texttt{boottest} \citep{RMNW}. It computes bootstrap confidence
intervals as well as bootstrap $P$ values, including ones for tests of
several linear restrictions on~$\bbeta$. In most cases, this is
inexpensive, even when $B$ is chosen to be a large number like 99,999;
see \citet*[Table~1]{MNW-guide}. However, computational cost can be an
issue when the number of regressors is very large.

%%%%%%%%%%%%%%%%%%%%%%%%%%%%%%%%%%%%%%%%%%%%%%%%%%%%%%%%
\section{Why Cluster-Robust Inference May Be Unreliable}
\label{sec:unreliable}
%%%%%%%%%%%%%%%%%%%%%%%%%%%%%%%%%%%%%%%%%%%%%%%%%%%%%%%%

None of the methods discussed in this paper yields exact inferences.
However, in many cases, some or all of them will yield inferences that
investigators believe to be sufficiently reliable. Consider a
hypothesis test at level~$\alpha$ or, equivalently, a confidence
interval at level $1-\alpha$. Let $\alpha_l\le\alpha$ and $\alpha_u\ge
\alpha$ be numbers that reflect an investigator's preferences for
reliability. For values of $\alpha_l$ and $\alpha_u$ sufficiently
close to~$\alpha$, most investigators would be happy if they could be
confident that the rejection frequency of a hypothesis test at
level~$\alpha$ were in the interval $[\alpha_l,\alpha_u]$ and that the
coverage of a confidence interval at level $1-\alpha$ were between
$1-\alpha_u$ and $1-\alpha_l$. As an example, it might be the case
that $\alpha_l=0.9\alpha$, and $\alpha_u=1.1\alpha$. Then, for tests
at the .05 level, an investigator would be prepared to tolerate
rejection frequencies of between 0.045 and~0.055. Of course, there is
no way to know precisely what these frequencies are for any
inferential procedure, but approximate methods are discussed in
\Cref{subsec:monte,subsec:placebo}.

It seems reasonable to call a procedure ``reliable'' if it can be
expected to yield a rejection frequency in the interval
$[\alpha_l,\alpha_u]$, or coverage in the interval
$[1-\alpha_u,1-\alpha_l]$, for values of $\alpha_l$ and $\alpha_u$
that an investigator is comfortable with. Different investigators may,
of course, choose different values. If an investigator chooses
$\alpha_l=0.999\alpha$ and $\alpha_u=1.001\alpha$, then there probably
exists no method for cluster-robust inference on samples of reasonable
size that is reliable. Even if such a method did exist, it would take
an enormous amount of computer time to verify that it achieves such a
high level of reliability. If an investigator insists on provably
conservative inference, then $\alpha_u=\alpha$ and $\alpha_l=0$, but
none of the methods discussed in this paper can achieve this.

It has been known for as long as cluster-robust inference has been
used that the number of clusters $G$ is far more important than the
number of observations $N$\tn. There are only $G$ independent score
vectors in \eqref{eq:betahat}. Thus the middle factor in expression
\eqref{eq:CV1} and its analogs for the other CRVEs involve summations
over $G$ terms, which are the outer products of the empirical score
vectors. Unless $G$ is large, random variation in these terms may
cause a CRVE to provide a poor approximation to the true variance
matrix \eqref{eq:trueV}. As $N/G$ increases, we might expect the
empirical scores (the $\hat\bis_g$) to provide better approximations
to the actual scores (the $\bis_g$), but the latter remain random. It
requires very strong assumptions to obtain exact results for $G$ fixed
as $N\to\infty$; see \citet*{BCH_2011} and \citet*{CSS_2021}.

In the early days of cluster-robust standard errors, it was commonly
conjectured that inference would be reliable whenever $G$ was
sufficiently large; the value $G=42$ is mentioned in
\citet*{MHE_2008}. This is a severe over-simplification. Although the
value of $G$ is certainly important, many other features of the data
can also greatly affect reliability.

For a CRVE to perform well, $1/G$ times the term $\sum_{g=1}^G
\hat\bis_g\hat\bis_g^\top$ in \eqref{eq:CV1}, and its analog for the
other CRVEs, must converge to the same matrix as $1/G$ times
$\sum_{g=1}^G \bis_g\bis_g^\top$. How well the former estimates the
latter certainly depends on $G$, but it also depends on how
homogeneous the clusters are; the less heterogeneity, the better. Any
sort of variation in the $\biX_g^\top\biX_g$ matrices and the
$\biX_g^\top\biy_g$ vectors can be damaging. Extreme variation in them
can lead to severely unreliable inferences.

Variation in the $N_g$, that is, the number of observations per
cluster, necessarily causes these quantities to vary. So does
variation in the distributions of the $\biX_g$ across clusters,
heteroskedasticity within and across clusters, and variation in the
patterns of within-cluster correlation. When the key regressor is a
treatment dummy, variation in treatment status across clusters can be
a particularly important source of heterogeneity. If only a few
clusters are treated, or not treated, then $(1/G)\sum_{g=1}^G
\hat\bis_g\hat\bis_g^\top$ may provide a very poor estimator of
$(1/G)\sum_{g=1}^G \bis_g\bis_g^\top$; see \citet{MW-JAE,MW-EJ}.

Several measures of cluster heterogeneity will be discussed in
\Cref{subsec:hetero}. When some of these measures suggest that there
is a lot of heterogeneity, many methods of cluster-robust inference
(perhaps all of them) are likely to be unreliable.

%%%%%%%%%%%%%%%%%%%%%%%%%%%%%%%%
\section{How Should We Cluster?}
\label{sec:how}
%%%%%%%%%%%%%%%%%%%%%%%%%%%%%%%%

Before we begin to compute standard errors, we need to decide just how
the scores are clustered. Sometimes there is more than one plausible
level of clustering. \Cref{subsec:level} deals with how to test a
finer level of clustering against a coarser one. Then
\Cref{subsec:twoway} briefly discusses two\tkk-way clustering. Not
surprisingly, results based on one\tkk-way clustering may be
completely untrustworthy if there is actually clustering in two (or
more) dimensions.

%%%%%%%%%%%%%%%%%%%%%%%%%%%%%%%%%%%%%%%%%%%%
\subsection{Testing the Level of Clustering}
\label{subsec:level}
%%%%%%%%%%%%%%%%%%%%%%%%%%%%%%%%%%%%%%%%%%%%

Almost all the literature on cluster-robust inference assumes that
observations are correctly assigned to clusters. If this assumption is
false, then inference is likely to be invalid. Of course, it is
impossible to test this assumption against all possible alternatives.
However, when different assignments of observations to clusters are
nested, there are ways to test for the correct level at which to
cluster.

Suppose that there are two levels at which it seems reasonable to
cluster. Clustering may either be fine (say, schools) or coarse (say,
school districts). If we mistakenly cluster at the fine level when
clustering is actually coarse, then cluster-robust inference will be
invalid. But if we mistakenly cluster at the coarse level when
clustering is actually fine, and the number of coarse clusters is not
large, inference is likely to be unreliable, as discussed in 
\Cref{sec:unreliable}. We will implicitly be relying on noisy
estimates of a lot of covariances that are actually zero. By
explicitly setting them to zero, fine clustering is likely to be much
more reliable.

\citet{IM_2016} proposes two tests for the level of clustering, both
of which involve re-estimating the model for various subsamples.
\citet{Cai-random} proposes a test based on randomization inference.
\citet*{MNW-testing} proposes both asymptotic and wild bootstrap tests
based on elements of the score vectors after regressors that are not
of primary interest have been partialed out. These are called
score\tkk-variance tests. They may be used to test the null of no
clustering against the alternative of any form of one\tkk-way
clustering, or the null of fine clustering against the alternative of
coarse clustering. Asymptotic score\tkk-variance tests are quite easy
to implement, but bootstrap ones can be computationally expensive. The
\texttt{Stata} package \texttt{mnwsvt} implements both of them.

Score\tkk-variance tests can be performed for one coefficient or
several. It is easiest to deal with a single coefficient, normally the
one of primary interest. Suppose this is the coefficient $\beta_j$ on
a regressor $\bix_j$. Let $\hat u_{ghi}$ denote the ordinary residual
for observation $i$ in fine cluster~$h$ within coarse cluster~$g$, and
let $z_{ghi}$ denote the residual from regressing $\bix_j$ on all the
other regressors. Then the empirical score for fine cluster $gh$ is
$\hat s_{gh} = \sum_i z_{ghi}\tk\hat u_{ghi}$. The numerator of the
score\tkk-variance test statistic is $\hat\theta = \hat\sigma_{\rm
c}^2 - \hat\sigma_{\rm f}^2$, where $\hat\sigma_{\rm c}^2$ is an
estimate of the variance of the scores under coarse clustering, and
$\hat\sigma_{\rm f}^2$ is an estimate of that variance under fine
clustering. Specifically, if $M_g$ is the number of fine clusters in
coarse cluster~$g$, then
\begin{equation}
\hat\sigma^2_{\rm c} = m_{\rm c} \sum_{g=1}^G 
\bigg(\sum_{h=1}^{M_g} \hat s_{gh}\!\bigg)^{\!\!2},
\quad\textrm{and}\quad
\hat\sigma^2_{\rm f} = m_{\rm f} \sum_{g=1}^G \sum_{h=1}^{M_g} 
\hat s_{gh}^2,
\end{equation}
where $m_{\rm c}$ and $m_{\rm f}$ are non-stochastic scaling factors.
Notice that $\hat\sigma^2_{\rm c}$ involves $\sum_{g=1}^G M_g^2$
terms, while $\hat\sigma^2_{\rm f}$ only involves $\sum_{g=1}^G M_g$
terms. When the null hypothesis of fine clustering is true, the terms
that appear in the former but not the latter tend to zero asymptotically.

If we divide $\hat\theta$ by an estimate of its standard deviation, we
obtain a test statistic that is asymptotically distributed as
$\N(0,1)$. This asymptotic test does not always perform well in finite
samples, and bootstrap tests generally appear to perform better. In
the package \texttt{mnwsvt}, the WCU-C bootstrap is used to generate
the bootstrap samples. However, the WCU-S bootstrap could be used
instead, and I conjecture that it would perform even better.

An important feature of score\tkk-variance tests is that they depend
on the coefficient(s) of interest. This is inevitable, because the
tests focus on the properties of the scores. For some coefficients, it
may be valid to cluster at the fine level, while for others it may
only be valid to cluster at the coarse level. Any decision we make
about the correct level at which to cluster has to depend on the
coefficient(s) that we care about. For this reason and others, it is
probably best to use score\tkk-variance tests as diagnostics rather
than as formal pre\tkk-tests.

%%%%%%%%%%%%%%%%%%%%%%%%%%%%%%%%%%%%%%%
\subsection{Two\tkk-\tn Way Clustering}
\label{subsec:twoway}
%%%%%%%%%%%%%%%%%%%%%%%%%%%%%%%%%%%%%%%

Up to this point, I have only discussed one\tkk-way clustering.
However, there are circumstances in which there may be two, or even
more than two, clustering dimensions. In the two\tkk-way case, every
observation is assumed to belong to one cluster in each of the two
dimensions. For example, there might be a spatial dimension and a time
dimension. If an observation belongs to cluster $g$ in the spatial
dimension and cluster $t$ in the time dimension, then it may be
correlated with other observations that belong either to cluster $g$
or to cluster $t$.

The idea of two\tkk-way clustering was independently discovered by
\citet{MH_2006}, \citet*{CGM_2011}, and \citet{Thompson_2011}. There
is a rapidly growing literature on two\tkk-way clustering, but much
less is known about its finite\tkk-sample properties than is the case
for one\tkk-way clustering. See, among others,
\citet*{MNW_2021,MNW-twoway}, \citet*{CHS_2024}, \citet{Vogel_2024},
\citet{Hounyo-boot}, and \citet*{Davezies_2025}.

Anything that causes inference to be unreliable for one\tkk-way
clustering also causes it to be unreliable for two\tkk-way clustering;
see \Cref{sec:unreliable}. If there are $G$ clusters in one dimension
and $H$ in the other, then neither $G$ nor $H$ should be too small,
and there should not be too much heterogeneity in either clustering
dimension. There are, however, some additional issues. For two\tkk-way
clustering, the filling in the sandwich for the true variance matrix is
\begin{equation}
\label{truesig}
\bSigma = \sum_{g=1}^G \bSigma_g + \sum_{h=1}^H \bSigma_h 
    - \sum_{g=1}^G\sum_{h=1}^H \bSigma_{gh}.
\end{equation}
Here, the $\bSigma_g$ and $\bSigma_h$ are the variance matrices of the
score vectors for each of the two clustering dimensions, and the
$\bSigma_{gh}$ are the variance matrices for the intersections of the
two dimensions. The third term has to be subtracted in order to avoid
double counting.

Although the matrix \eqref{truesig} is positive definite under
reasonable assumptions, its empirical analogs are not. The simplest
empirical analog involves replacing $\bSigma_g$ by
$\hat\bis_g\hat\bis_g^\top$, and similarly for $\bSigma_h$ and
$\bSigma_{gh}$, but there is a better one based on the
cluster-jackknife estimator \eqref{eq:CV3}; see \citet*{MNW-twoway}.
When $\hat\bSigma$, an estimator of $\bSigma$, is not positive
definite, standard errors may be undefined. Even more worrisome, when
$\hat\bSigma$ is close to but not quite singular, standard errors may
be positive but extremely small.

Two\tkk-way clustered standard errors can never be smaller,
asymptotically, than one\tkk-way clustered ones in either dimension,
although they might be the same if the relevant scores were only
clustered in one dimension. Thus if a two\tkk-way standard error is
smaller than either of the one\tkk-way ones, it must be so because of
sampling variation and probably should not be believed. To get around
this problem, \citet*{MNW-twoway} and \citet*{Davezies_2025}
independently proposed computing standard errors based on both
dimensions of one\tkk-way clustering along with two\tkk-way ones, when
those are defined, and using whichever is largest. Especially when
CV$_{\tn3}$ standard errors are used, this procedure seems to work
quite well when there really is two\tkk-way clustering.

It is tempting to think that the empirical analog of the last matrix
on the right-hand side of \eqref{truesig}, without the minus sign,
should be used instead of the empirical analog of \eqref{truesig}
itself. This would be valid if the scores were correlated only within
the intersections instead of within each of the two dimensions. In
most cases, however, such an assumption is highly implausible. If it
were true, then we should obtain roughly the same standard errors
whether we cluster by the first dimension, the second dimension, or
their intersection.

Unfortunately, there is at present no formal way to test one\tkk-way
clustering against two\tkk-way clustering, although
\citet*[Remark~9]{MNW-testing} conjectures that a version of the
score\tkk-variance test can be used. In practice, there are often
compelling reasons to cluster in one dimension and much less
compelling ones to cluster in a second dimension. In such cases, it
probably makes sense to stick with one\tkk-way clustering in the first
dimension unless two\tkk-way clustering leads to noticeably larger
standard errors.

%%%%%%%%%%%%%%%%%%%%%%%%%%%%%%%%%%%%%%%%%%%
\section{Methods for Assessing Reliability}
\label{sec:assess}
%%%%%%%%%%%%%%%%%%%%%%%%%%%%%%%%%%%%%%%%%%%

There are several diagnostic measures that do not directly estimate
the rejection frequencies or coverage probabilities associated with
any particular method but can often provide valuable information.
These measures, discussed in \Cref{subsec:hetero}, can raise red flags
which suggest that some inferential methods are unlikely to be
reliable for the model and sample of interest. When there are no red
flags, most methods should perform well.

The other methods discussed in this section do directly estimate how
well alternative methods perform. \Cref{subsec:monte} deals with ways
to estimate rejection frequencies (or, equivalently, coverage) for a
specific model and dataset by performing Monte Carlo experiments
targeted at that case. \Cref{subsec:placebo} then discusses placebo
regressions. Targeted Monte Carlo experiments and placebo regressions
both involve performing a large number of simulations, so they can be
computationally expensive. They each have advantages and
disadvantages, but if they yield similar results, then those results
can probably be relied upon.

%%%%%%%%%%%%%%%%%%%%%%%%%%%%%%%%%%%%%%%%%%%%%%
\subsection{Measures of Cluster Heterogeneity}
\label{subsec:hetero}
%%%%%%%%%%%%%%%%%%%%%%%%%%%%%%%%%%%%%%%%%%%%%%

The smaller the number of clusters $G$, the less reliable all types of
cluster-robust inference are likely to be. Unfortunately, there is no
simple rule about how large $G$ needs to be. It depends in complicated
ways on many features of the model and data. For a given $G$, the more
the data vary across clusters, the less reliable inference tends to
be.

The most obvious way in which clusters can differ is in how large they
are. When the $N_g$ vary greatly, we usually cannot expect any method
to work really well. Serious problems tend to occur when there are a
few clusters that are excessively large relative to the average.
However, they do not occur when there are a few clusters that are
excessively small. Imagine two samples, each with 20 clusters. In both
of them, each of the first 19 clusters has 500 observations. But in
the first sample, cluster number 20 has 10 observations, while in the
second sample it has 10,000 observations. If appropriate methods are
used, inferences from the first sample may well be quite reliable,
since it is almost like a sample with 19 equal-sized clusters; we
should perhaps use the $t(G-2)$ distribution instead of $t(G-1)$. In
contrast, inferences from the second sample are likely to be very
unreliable for all methods, because a single cluster contains more
than half the observations. Monte Carlo evidence on such extreme cases
may be found in \citet*[Figure~3]{DMN_2019} and
\citet*[Figure~7]{MNW-bootknife}.

One way to deal with samples that include a small number of very large
clusters is to use weighted least squares (WLS) instead of OLS. Such a
method is proposed in \citet*[Section~5]{CSW_2025}, where both the
$y_{gi}$ and $\biX_{gi}$ are multiplied by $N_g^{-1/2}$ so that every
cluster is given the same weight. For the second example above, this
might well be a good thing to do. For the first example, however, it
would surely yield inefficient estimates, because the 10 observations
in the last cluster would be weighted much more heavily than any of
the other observations. There are undoubtedly cases in which
down-weighting large clusters would be desirable, but not much is
currently known about this sort of procedure. Applying any sort of WLS
in a mechanical fashion is probably not a good idea at present.

As was noted in \Cref{sec:unreliable}, variation in the $N_g$ is not
the only source of heterogeneity across clusters. Whatever the source,
large variation in the $\biX_g^\top\biX_g$ matrices and the
$\biX_g^\top\biy_g$ vectors tends to be associated with unreliable
inference. In particular, \citet*{MNW-influence} shows that inference
tends to become less reliable as the leverage of different clusters
becomes more variable.

The concept of leverage \citep*{BKW_1980} is normally applied at the
observation level, but it can just as well be applied at the cluster
level. There are at least two measures of cluster-level leverage. The
one that seems to be most associated with unreliable inference is
partial leverage \citep{CW_1980}, which can be calculated separately
for each coefficient. Suppose that we are interested in the \th{j}
regressor, contained in the vector $\bix_j$. If we regress $\bix_j$ on
all the other regressors, yielding the vector of residuals
$\acute\bix_j$, the partial leverage for regressor $j$ for the \th{g}
cluster is
\begin{equation}
L_{gj} = \frac{\acute\bix_{gj}^\top\acute\bix_{gj}}
{\acute\bix_j^\top\acute\bix_j}\tk,
\label{eq:partial}
\end{equation}
where $\acute\bix_{gj}$ is the subvector of $\acute\bix_j$
corresponding to the \th{g} cluster. It is easy to calculate 
\eqref{eq:partial} for every cluster for any coefficient of interest.
Since the $L_{gj}$ sum to unity, their average is evidently $1/G$.
Thus, if cluster $h$ has $L_{hj} >\!> 1/G$, it has high partial
leverage for the \th{j} coefficient.

When $G$ is small, it can be illuminating to look at all the $L_{gj}$.
When $G$ is not small, it is probably better to graph them or to
report summary measures of how much they vary across clusters. One
such measure is the scaled variance
\begin{equation}
V^j_s = \frac{G^2}{(G-1)} \sum_{g=1}^G  (L_{gj} - 1/G)^2.
\label{eq:coefvar}
\end{equation}
This will be zero whenever every cluster has the same partial
leverage, and it will be large whenever the variance of the $L_{gj}$
is large relative to $1/G^{\tk2}$, the square of their mean. The square
root of $V^j_s$ is the coefficient of variation of the partial
leverages.

\Citet*{CSS_2017} proposes a family of measures $G_j^*(\rho)$ called
the ``effective number of clusters.'' These measures depend on a
parameter $\rho$, which would be the intra-cluster correlation for the
disturbances in a random-effects model (see \Cref{subsec:monte}), and
they vary by coefficient. The original paper suggests using $\rho=1$.
However, when there are cluster fixed effects, the only feasible value
of~$\rho$ is~0, because otherwise the fixed effects cause there to be
singularities in the computation of $G_j^*(\rho)$.

It can be shown that $G^*_j(0)$ is simply a monotonically decreasing
function of the scaled variance defined in \eqref{eq:coefvar}
\citep*[Section~2.3]{MNW-influence}. Thus, when $V^j_s$ is large,
$G^*_j(0)$ is necessarily much smaller than $G$. Although $V^j_s$ and
$G^*_j(0)$ convey exactly the same information, the latter is easier
to interpret, because it is bounded above by~$G$. If, for example,
$G=50$ and $G^*_j(0)=11.2$, we would expect cluster-robust inference
to be challenging. On the other hand, if $G^*(0)=48.7$, we would
expect many methods to yield reasonably reliable inferences.

Many regression models focus on a treatment dummy variable which
varies at the cluster level, or, in the case of
difference-in-differences (DiD) models, at the cluster-time level. For
clusters that are never treated, the treatment dummy equals~0; for
clusters that are always treated, it equals~1; and for clusters that
are sometimes treated, it equals either~0 or~1. When treatment varies
at the cluster level, the sample must contain both treated and control
clusters if the treatment effect is to be identified. For reliable
inference, such a sample must contain enough clusters of each type. If
$G_1$ clusters are treated and $G_0$ are controls, then neither $G_1$
nor $G_0=G-G_1$ should be too small. For DiD models, it is also
important that $G_1$ not be too small, but $G_0$ does not have to be
positive if there are enough sometimes\tkk-treated clusters.

When every cluster is either treated or not, it is easy to see that
having very few treated or control clusters can lead to extreme
cluster heterogeneity and, consequently, very unreliable inference.
The few treated (or control) clusters necessarily have high leverage,
because omitting any of them could potentially change the estimates
and standard errors a lot. When there is just one treated or one
control cluster, omitting it makes it impossible to estimate the
treatment coefficient, so that CV$_{\tn2}$ and CV$_{\tn3}$ standard
errors cannot even be computed. In this case, CV$_{\tn1}$ standard
errors can be computed, but they are usually far smaller than they
should be.

To what extent inference is reliable when the number of treated (or
control) clusters is small depends in a complicated way on $G$, $G_1$,
the sizes of the treated and control clusters, the distributions of
the other regressors, and the distributions of the disturbances.
Simulation evidence \citep*{MNW-bootknife,Hansen_2025} suggests that
no methods are always reliable in such cases, but that some methods
can work much better than others. The problem is not simply
over-rejection. Some methods, including all variants of the restricted
wild cluster bootstrap, actually tend to under-reject, sometimes
severely. \citet{MW-JAE,MW-EJ} provide an algebraic explanation of
this phenomenon.

The \texttt{Stata} package \texttt{summclust} \citep*{MNW-influence}
calculates both leverage and partial leverage at the cluster level, as
well as measures of how much each of them varies, the effective number
of clusters, and several other useful diagnostics. It also calculates
both CV$_{\tn1}$ and CV$_{\tn3}$ variance matrices, along with the
associated $P$ values and confidence intervals.

None of the diagnostics discussed so far depends in any way on the
regressand $\biy$. However, the way in which the disturbances are
distributed can matter substantially for the reliability of inference.
In particular, there is evidence that both the extent of
heteroskedasticity across clusters, if any, and the patterns of
intra-cluster correlation can be important. For the former, it is easy
to calculate the average variance of the residuals for each cluster,
\begin{equation}
\hat\sigma_g^2 = \frac1{N_g-1} \sum_{i=1}^{N_g} (\hat u_{gi} - \bar
u_g)^2, \quad g=1,\ldots,G,
\label{clustvar}
\end{equation}
where $\bar u_g$ is the average residual for cluster~$g$. When there
are cluster fixed effects, $\bar u_g=0$ for all~$g$. In this case, we
can compute all of the $\hat\sigma_g^2$ at once by regressing the
squared residuals on $G$ cluster dummies, although this will
implicitly divide by $N_g$ instead of $N_g-1$. When the
$\hat\sigma_g^2$ vary substantially across clusters, inference is
likely to be unreliable; see \citet[Online Appendix]{MW-EJ} and
\citet*[Figure~5]{DMN_2019}.

The Monte Carlo simulations in a few recent papers
\citep*{Hansen_2025,CSW_2025} employ DGPs where $\var(u_{gi})$ is
always much larger for treated observations than for controls. Thus,
when the null is true, the treatment is assumed to affect the variance
of the dependent variable but not its mean, a specification that seems
rather hard to justify. When the data are generated in this way, all
tests and confidence intervals seem to be less reliable than they
would otherwise be, especially ones based on the wild cluster
bootstrap. If an investigator believes that the data may display this
sort of heteroskedasticity, then it is easy to check for it. They
simply need to run the regression
\begin{equation}
\hat u_{gi}^2 = \eta_1 + \eta_2\tk T_{gi} + e_{gi}, \quad g=1,\ldots,G,
\quad i=1,\ldots,N_g.
\label{twovars}
\end{equation}
where $T_{gi}$ is the treatment dummy for observation~$gi$. In
regression \eqref{twovars}, $\eta_1$ is the average variance for
control observations, and $\eta_1 + \eta_2$ is the average variance
for treated observations. A cluster-robust test for $\eta_2=0$ can be
used to test whether the disturbances for control and treated clusters
have the same average variance. If $\hat\eta_2$ is positive,
significant, and not small relative to $\hat\eta_1$, then results from
inferential methods that might otherwise be expected to perform well
should probably be interpreted with caution.

Another diagnostic that depends on both $\biy$ and $\biX$ is the
variability in the $\hat\bbeta^{(g)}$, the omit-\tkk one\tkk-cluster
estimates defined in \eqref{eq:delone}. Since these are needed to
compute CV$_{\tn3}$, it is almost costless to report them, and
\texttt{summclust} optionally does so. When there are one or two
clusters where $\hat\bbeta^{(g)}$ differs greatly from $\hat\bbeta$,
the investigator should be cautious. Such a situation not only
suggests that inferences may be unreliable, but also calls into
question whether the model is valid for every cluster.

%%%%%%%%%%%%%%%%%%%%%%%%%%%%%%%%%%%%%%%%%%%%%
\subsection{Targeted Monte Carlo Experiments}
\label{subsec:monte}
%%%%%%%%%%%%%%%%%%%%%%%%%%%%%%%%%%%%%%%%%%%%%

The methods discussed in \Cref{subsec:hetero} provide qualitative
information about how well various types of cluster-robust inference
can be expected to perform for a particular model and dataset.
However, they do not provide quantitative estimates of how accurate
$P$ values and confidence intervals are likely to be. There are at
least two ways to do so. Neither can be expected to work perfectly,
but limited experience suggests that both can work quite well.

The first approach is to perform a Monte Carlo experiment targeted at
the model and dataset of interest. Over the years, a great many Monte
Carlo experiments have been devoted to studying the performance of
various procedures for cluster-robust inference. Most of these,
however, are not designed to provide guidance for any particular
empirical application. Instead, designers of experiments typically
choose $G$, $N$\tn, the $N_g$, the number of regressors and how they
are distributed, the distributions of the $u_{gi}$, and so on, for a
variety of reasons, including ease of programming and minimizing
computing time. The objective is often to study how changes in some
aspects of the experimental design affect the performance of
alternative procedures. Although this can yield valuable information,
many simulated datasets probably do not look much like actual ones.

Nevertheless, a good deal has been learned from traditional Monte
Carlo experiments. In particular, many experiments suggest that, if
$G$ is reasonably large and the clusters do not vary much in size or
leverage, the better inferential methods will often yield very similar
results. Thus, when this is the case for a particular empirical model,
there is probably no need for the investigator to worry about the
reliability of those results. In cases where $G$ is not large,
however, or clusters vary a lot in size or leverage, or there are
other complications such as heteroskedasticity, existing simulation
results may not provide much guidance. Unless all the features of a
particular empirical application closely resemble those of some
published simulation design, it is not really possible for an
empirical investigator to decide with any confidence which inferential
procedures should be relied upon.

In contrast, a Monte Carlo experiment targeted at a particular model
and dataset can help an investigator to make that decision. In most
respects, performing a targeted Monte Carlo experiment is quite easy.
In order to generate a dataset for each of $R$ replications, we simply
use the actual matrix $\biX$ and the actual clusters, along with a
vector $\bbeta$ that we specify. It could equal either $\bzero$ or
$\hat\bbeta$, because, when the DGP is \eqref{eq:lrmodel}, the value
of $\bbeta$ merely affects what hypothesis should be tested and the
locations, but not the lengths, of confidence intervals. Since $\biX$
is the same for every replication, so are all the $\biX_g^\top\biX_g$,
which means that computation should be inexpensive.

The main difficulty with targeted Monte Carlo experiments is that the
investigator has to decide precisely how to generate the~$\biu_g$ in
order to generate the vector $\biy$ for each simulated sample. The
easiest approach is to use the random-effects model
\begin{equation}
u_{gi} = v_g + \epsilon_{gi}, \quad v_g\sim\N(0,\sigma_v^2), \quad
\epsilon_{gi}\sim\N(0,\sigma_\epsilon^2),
\label{eq:RE}
\end{equation}
which implies that the disturbances within each cluster are
equi-correlated. The values of $\sigma_v^2$ and $\sigma_\epsilon^2$
are important, since they determine the value of the equi-correlation
coefficient, $\rho=\sigma_v^2/(\sigma_v^2+\sigma_\epsilon^2)$. Many
published Monte Carlo experiments have used \eqref{eq:RE}. However, if
the regressors include cluster fixed effects, the latter completely
explain the $v_g$, and there is no longer any within-cluster
correlation. Thus, when the random-effects specification \eqref{eq:RE}
is used for a model with fixed effects, there is no need to employ
cluster-robust inference at all. Simply using a
heteroskedasticity-robust estimator like HC$_2$ or HC$_3$ is
asymptotically valid and will probably work better than using a CRVE,
at least when $G$ is not large.

It is not at all clear how best to generate the disturbances for
models with cluster fixed effects. Various methods have been used in
\citet*{MNW-bootknife,MNW-testing,MNW-logit}, \citet{Hansen_2025}, and
other papers. Precisely how the disturbances are generated can matter.
When one is performing a targeted Monte Carlo experiment, it is
probably a good idea to try several different specifications, with
varying patterns of intra-cluster correlation, to see whether they
lead to essentially the same results.

%%%%%%%%%%%%%%%%%%%%%%%%%%%%%%%%
\subsection{Placebo Regressions}
\label{subsec:placebo}
%%%%%%%%%%%%%%%%%%%%%%%%%%%%%%%%

A different sort of simulation experiment is based on the idea of
placebo regressions \citep*{BDM_2004}. Once again, we estimate the
model $R$ times. But instead of holding $\biX$ fixed and generating
the vector $\biy$ for each replication, we hold $\biy$ fixed and vary
one column of the $\biX$ matrix. The column that varies, say $\biz$,
is a ``placebo regressor,'' a completely artificial regressor that
should not be related to $\biy$ at all. The idea is to
construct~$\biz$ at random so that it resembles the regressor we are
really interested in. Just how to do this will vary from case to case
and may require some thought. See, among others, \citet{CT-persist}
and \citet*{MNW-logit}.

For example, suppose that a treatment regressor varies at the cluster
level and equals~1 for 20 clusters and~0 for 22 clusters. Then it
seems natural to generate $\biz$ for each simulated sample so that it
equals~1 for 20 randomly chosen clusters and~0 for the remaining ones.
In this case, since there are $_{42}C_{20} = 5.14\times 10^{11}$
random ways to choose $\biz$, it is surely safe to generate the
elements of $\biz$ with replacement. When the number of possible 
vectors $\biz$ is not so large, it is desirable to generate them
without replacement in order to avoid duplicates.

The placebo regression approach is not feasible if there are only a
few ways to generate the placebo regressors. For example, if there
were ten clusters, only one of which was treated, there would be only
nine possible ways to generate placebo regressors that were not
identical to the actual treatment regressor. Unfortunately, we cannot
hope to learn very much from an empirical distribution based on just
nine points. If there were a moderate number of possible placebo
regressors, perhaps between 100 and a few hundred thousand, then we
certainly could learn something. However, to avoid simulation error in
such cases, it would be better to enumerate all the possible placebo
regressors rather than choosing them at random with or without
replacement.

The advantage of performing a placebo regression experiment instead of
a targeted Monte Carlo experiment is that the former preserves
whatever process generated the disturbances for the actual dataset.
However, there are some disadvantages. Just how the $\biz$ are
generated can matter. There is often more than one way to do so, and
$\biz$ may or may not have the same features as $\bix_j$, the
regressor that is really of interest. Moreover, it is not entirely
clear how to specify the placebo regressions.

We can rewrite model \eqref{eq:lrmodel} for the entire sample as
\begin{equation}
\label{eq:placebo} 
\biy =\biX_1\bbeta_1 + \beta_j\tk\bix_j + \biu,
\end{equation}
where $\biX = [\biX_1\;\;\bix_j]$. The coefficient of interest is
$\beta_j$. Given some procedure for generating the $\biz$ vectors,
there are now two ways to proceed. The simplest one is to replace
$\bix_j$ in \eqref{eq:placebo} by a different realization of $\biz$
for each of the $R$ placebo regressions. This is what a number of
authors have done; see, for example, \citet{CT-persist}. However, this
may not always be a good idea. If $\bix_j$ actually belongs in the
DGP, then omitting it would cause the model to be misspecified. In
that case, a test of the hypothesis that the coefficient on~$\biz$
is~0 may over-reject, not because an inferential method is unreliable,
but simply because the placebo regressor~$\biz$ is being added to a
misspecified model \citep{DM-interp}. Whether and to what extent this
happens will depend on the value of $\beta_j$ and the correlation
between $\biM_1\bix_j$ and $\biM_1\biz$, where $\biM_1 = \bfI -
\biX_1(\biX_1^\top\biX_1)^{-1}\biX_1^\top$.

This suggests that we may want to use a second approach, which may be
safer. It is to include $\bix_j$ along with $\biz$ in the placebo
regressions. However, this means that they will have one more
regressor than the actual regression \eqref{eq:placebo}, which is not
entirely satisfactory. Both approaches are illustrated in
\Cref{sec:applic}.

As just described, placebo regressions cannot handle
heteroskedasticity related to the test regressor, because $\biy$ does
not vary across the simulations. If this sort of heteroskedasticity
exists (see \eqref{twovars} and the discussion around it), then the
variances of the $y_{gi}$ would have to depend on the $z_{gi}$, that
is, the values of the placebo regressor. But doing this would seem to
violate the spirit of placebo regressions, and it is not at all clear
how to modify the $y_{gi}$ in such a way. In contrast, there is no
difficulty incorporating any hypothesized form of heteroskedasticity
into targeted Monte Carlo experiments.

%%%%%%%%%%%%%%%%%%%%%%%%%%%%%%%%%%%%
\section{Two Empirical Applications}
\label{sec:applic}
%%%%%%%%%%%%%%%%%%%%%%%%%%%%%%%%%%%%

In this section, the methods discussed in \Cref{sec:assess} are
applied to two empirical applications. Both applications involve
binary dependent variables in educational settings where some students
are treated and others are not. However, there are some interesting
differences between the two applications.

%%%%%%%%%%%%%%%%%%%%%%%%%%%%%%%%%%%%%%%%%%%%
\subsection{Female Role Models in Economics}
\label{subsec:role}
%%%%%%%%%%%%%%%%%%%%%%%%%%%%%%%%%%%%%%%%%%%%

\citet{PS_2020} studies a field experiment designed to investigate
whether exposing female students in first-year economics courses to
``successful and charismatic women who majored in economics at the
same university'' makes them more likely to major in economics or take
further economics courses. I focus on the last column of their
Table~3, where the regressand is a binary variable that equals~1 if a
student took at least one more economics course. Each observation
corresponds to a female student. There are 627 observations in 12
classes which were taught in 2015, 2016, or both years. Of these, 130
students were ``treated'' in 2016. All the treated students were in 4
of the 12 classes. The original paper only reports OLS results for a
linear probability model (LPM), but I also estimate a logit model.

\begin{table}[tp]
\caption{Effects of Treatment on Taking Another Economics Course}
\label{tab:another}
\vspace*{-12pt}
\begin{center}
\begin{tabular*}{0.98\textwidth}{@{\extracolsep{\fill}}
lcccccc}
\toprule
Method &\textrm{Coef.} &\textrm{Std.\ error}
&\textrm{$t$-statistic} &\textrm{$P$ value} &\textrm{CI Lower}
&\textrm{CI Upper}\\
\midrule
LPM HC$_1$ &0.1389 &0.0673 &2.0632 &0.0395 &$\phm0.0067$ &0.2710 \\
LPM CV$_1$ &0.1389 &0.0518 &2.6791 &0.0214 &$\phm0.0248$ &0.2529 \\
LPM CV$_3$ &0.1389 &0.0646 &2.1505 &0.0546 &$-0.0033$ &0.2810 \\
LPM CV$_{\tn3}^{\rm BH}$ + $t({\rm BH})$ &0.1389 &0.0674 &2.0589 &0.0504
  &$-0.0004$ &0.2781 \\
LPM Pairs &0.1389 & &2.6791 &0.1019 &$-0.0087$ &0.3796 \\
LPM Pairs (boot s.e.) &0.1389 &0.0601 &2.3108 &0.0412 &$\phm0.0066$ &0.2711 \\
LPM WCU-C &0.1389 & &2.6791 &0.0328 &$\phm0.1037$ &0.2670 \\
LPM WCU-C (boot s.e.) &0.1389 &0.0492 &2.8214 &0.0166 &$\phm0.0305$ &0.2472 \\
LPM WCU-S &0.1389 & &2.6791 &0.0440 &$\phm0.0342$ &0.2740 \\
LPM WCU-S (boot s.e.) &0.1389 &0.0674 &2.0590 &0.0640 &$-0.0096$ &0.2873 \\
LPM WCR-C &0.1389 & &2.6791 &0.0344 &$\phm0.0133$ &0.2617 \\
LPM WCR-S &0.1389 & &2.6791 &0.0402 &$\phm0.0079$ &0.2573 \\
\midrule
Logit (default) &0.8739 &0.4071 &2.1467 &0.0322 &$\phm0.0743$ &1.6733 \\
Logit CV$_{\tn1}$ &0.8739 &0.3112 &2.8079 &0.0170
  &$\phm0.1889$ &1.5589 \\
Logit CV$_{\tn3{\rm L}}$ &0.8739 &0.3875 &2.2554 &0.0455
  &$\phm0.0211$ &1.7266 \\
Logit WCLU-C &0.8739 & &2.8079 &0.0111 &$\phm0.1736$ &1.5728 \\
Logit WCLU-C (boot s.e.) &0.8739 &0.2955 &2.9575 &0.0130 &$\phm0.2236$
  &1.5242 \\
Logit WCLU-S &0.8739 & &2.8079 &0.0211 &$\phm0.1242$ &1.6223 \\
Logit WCLU-S (boot s.e.) &0.8739 &0.4045 &2.1602 &0.0537 &$-0.0166$ &1.7644 \\
Logit WCLR-C &0.8739 & &2.8079 &0.0296 \\
Logit WCLR-S &0.8739 & &2.8079 &0.0345 \\
\bottomrule
\end{tabular*}
\end{center}
\textbf{Notes:}\par
$\bullet$ There are 999,999 bootstrap samples. WC bootstraps
use six-point weights \citep{Webb_six}.\par
$\bullet$ ``CI Lower'' and ``CI Upper'' are limits of 95\% confidence
intervals.\par
$\bullet$ For CV$_{\tn3}^{\rm BH}$ + $t({\rm BH})$, the CV$_{\tn3}$
standard error and the degrees of freedom for the $t$ distribution are
calculated using the procedure of \citet{Hansen-jack}.\par
$\bullet$ For the logit model, CV$_{\tn3{\rm L}}$ is a cluster-jackknife
standard error based on the linearization proposed in \citet*{MNW-logit}.\par
$\bullet$ Unrestricted bootstrap methods marked ``boot s.e.'' use
bootstrap standard errors. The others report $P$ values for
CV$_{\tn1}$ $t$-statistics and studentized bootstrap confidence
intervals.
\end{table}

This is an application for which cluster-robust inference would not
generally be expected to work well. The paper clusters by class, so
there are just 12 clusters and 4 treated clusters. There are no fixed
effects, since they would explain the treatment dummy. Because cluster
sizes vary between 12 and 104, some clusters have much more leverage
than others. In consequence, the values of $G^*(0)$ and $G^*(1)$ are
just 8.49 and 5.98, respectively. 

\Cref{tab:another} shows $t$-statistics, $P$ values, and confidence
intervals for a large number of methods. Some results are surprising.
In particular, the CV$_{\tn1}$ standard errors for both the LPM and
the logit model are substantially smaller than
heteroskedasticity-robust, or default logit, standard errors that do
not allow for clustering. This suggests that the CV$_{\tn1}$ standard
errors are biased downwards, which would not be at all surprising in
view of the small numbers of clusters and treated clusters. Note that
the original paper does not rely directly on the CV$_{\tn1}$ standard
error for the LPM. Instead, it reports a bootstrap $P$ value of 0.032
based on the WCR-C bootstrap with $B=1000$, which differs slightly
from the corresponding value of 0.0345 in the table, almost certainly
just because of simulation randomness.

Most of the results in \Cref{tab:another} suggest that the $P$ value
for the treatment dummy is less than 0.05, but there are some
exceptions. For the LPM, the CV$_{\tn3}$ standard error when either
combined with $t(11)$ critical values or subjected to the procedures
of \citet{Hansen-jack} yields $P$ values slightly greater than 0.05.
The pairs cluster bootstrap yields two contradictory results. The
bootstrap $P$ value \eqref{symbootP} is greater than 0.10, but the $P$
value for a $t$-statistic based on the bootstrap standard error
is~0.041. The contradiction goes in the other direction for the WCU-S
and WCLU-S bootstraps. Their bootstrap $P$ values are both less than
0.05, but the ones for their bootstrap $t$-statistics are both greater
than 0.05.

For the LPM and logit models, WCU-S and \mbox{WCLU-S} standard errors
yield $t$-statistics of 2.0590 and 2.1602, respectively. These differ
greatly from the ones for WCU-C and WCLU-C but are extremely close to
the ones without clustering. Note that the table does not report
standard errors for procedures that simply bootstrap the CV$_{\tn1}$
$t$-statistic.

In order to see which results in \Cref{tab:another}, if any, can be
relied upon, I performed eleven Monte Carlo experiments and two
placebo\tkk-regression experiments. For the former, the data are
generated by a linear probability model estimated by OLS subject to
the restriction that the coefficient on the treatment dummy is zero.
This yields fitted values $\biX_{gi}\tilde\bbeta$. The disturbances
are then generated by the random-effects model \eqref{eq:RE}, with the
two variance parameters chosen so that $\var(u_{gi})=1$. The
coefficient of intra-cluster correlation, $\rho$, varies from 0.00 to
0.50 by increments of~0.05. Then $y_{gi}=0$ when $\Phi(u_{gi}) \ge
\biX_{gi}\tilde\bbeta$, and $y_{gi}=1$ otherwise, where $\Phi(\cdot)$
is the cumulative normal distribution function. This procedure yields
average values of $y_{gi}$ extremely close to the observed value
of~0.217.

\begin{figure}[tb]
\begin{center}
\caption{Monte Carlo rejection frequencies as functions of $\rho$}
\label{fig:A}
\includegraphics[width=0.95\textwidth]{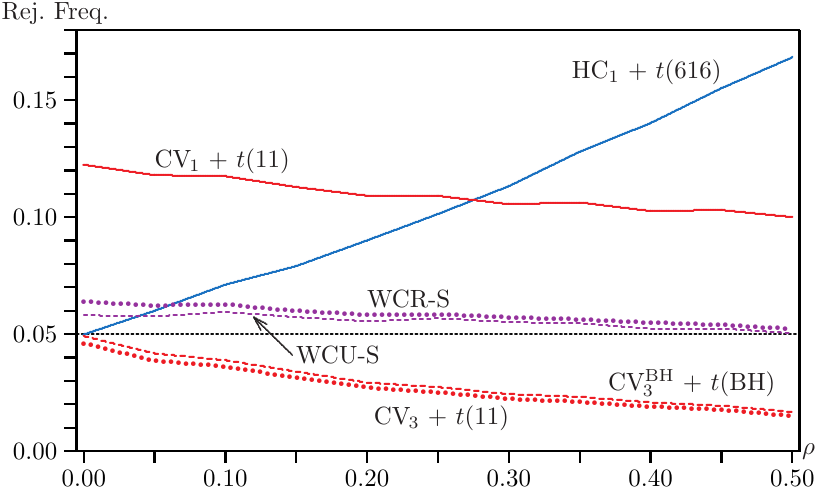}
\end{center}
\textbf{Notes:}\par
$\bullet$ Each Monte Carlo experiment uses 100,000 replications.\par
$\bullet$ Bootstrap tests use 999 bootstrap samples and Webb's
six-point distribution.
\end{figure}

For the Monte Carlo experiments, the value of $\rho$ matters greatly;
see \Cref{fig:A}. When $\rho=0$, HC$_1$ yields essentially perfect
inferences, while CV$_{\tn1}$ combined with $t(11)$ over-rejects
substantially, CV$_{\tn3}$ under-rejects slightly, and Hansen's
procedure under-rejects a little bit less. The WCR-S and WCU-S
bootstraps both over-reject somewhat. Oddly, the former does so to a
greater extent than the latter. As $\rho$ increases, HC$_1$
over-rejects more and more severely, while all other methods reject
less frequently. Except for very small values of~$\rho$, where
Hansen's procedure performs particularly well, the two bootstrap
methods seem to perform the best.

The increasing under-rejection by the jackknife methods as $\rho$
increases that is evident in \Cref{fig:A} is surprising. What seems to
be happening is that the correlation between the absolute value of the
treatment coefficient and the CV$_{\tn3}$ standard error increases
steadily as~$\rho$ increases, from $-0.008$ when $\rho=0.0$ to 0.205
when $\rho=0.5$. Thus the distribution of the CV$_{\tn3}$
$t$-statistic becomes less spread out as~$\rho$ increases, although it
is always being compared with the $t(11)$ distribution. The same thing
happens for the CV$_{\tn1}$ $t$-statistic, but because the standard
error is always much too small, there is over-rejection even for large
values of~$\rho$.

As discussed in \Cref{subsec:placebo}, there are two ways to handle
the placebo regressor. Thus there are two placebo\tkk-regression
experiments.  In the Placebo\tkk-A experiment (``A'' is for ``add''),
the placebo regressor is an additional regressor. In the Placebo\tkk-R
experiment (``R'' is for ``replace''), it replaces the actual
treatment regressor. In both experiments, the placebo regressor looks
just like the actual treatment regressor, except that treatment is
assigned at random to observations within each of the four treated
clusters. This was done with replacement. Since cluster~1, for
example, has 78 observations, of which 33 are treated, there are
approximately $10^{22}$ ways to assign treatment to observations
within that cluster.

\begin{table}[tp]
\caption{$P$ Values and Simulated Rejection Frequencies for
Role-Model Example}
\label{tab:prf}
\vspace*{-12pt}
\begin{center}
\begin{tabular*}{0.98\textwidth}{@{\extracolsep{\fill}}
lccccc}
\toprule
Method &\textrm{$P$ value} &MC (\tk$\rho=0$) &MC (\tk$\rho=0.25$)
&Placebo\tkk-A &Placebo\tkk-R \\
\midrule
HC$_1$ + $t(616)$ &0.0395 &0.0495 &0.1010 &0.0533 &0.0519 \\
CV$_{\tn1} + t(11)$ &0.0214 &0.1222 &0.1088 &0.1872 &0.1867 \\
CV$_{\tn3} + t(11)$ &0.0546 &0.0454 &0.0245 &0.0936 &0.0933 \\
CV$_{\tn3}^{\rm BH} + t({\rm BH})$ &0.0504 &0.0486 &0.0269 &0.0461 &0.0459 \\
WCU-S bootstrap &0.0440 &0.0576 &0.0563 &0.0764 &0.0762 \\
WCR-S bootstrap &0.0402 &0.0632 &0.0578 &0.0634 &0.0643 \\
Pairs bootstrap &0.1019 &0.0247 &0.0145 &0.0242 &0.0272 \\
\midrule
Logit ML + $t(616)$ &0.0322 &0.0503 &0.1036 &0.0467 &0.0449 \\
Logit CV$_{\tn1} + t(11)$ &0.0170 &0.0873 &0.0825 &0.1486 &0.1480 \\
Logit CV$_{\tn3{\rm L}} + t(11)$ &0.0455 &0.0427 &0.0264 &0.0897 &0.0893 \\
Logit WCLU-S &0.0211 &0.0568 &0.0643 &0.0763 &0.0765 \\
Logit WCLR-S &0.0345 &0.0617 &0.0635 &0.0574 &0.0587 \\
\bottomrule
\end{tabular*}
\end{center}
\textbf{Notes:}\par
$\bullet$ The column headed ``$P$ value'' contains analytic or bootstrap
$P$ values, taken from \Cref{tab:another}. The other columns contain
rejection frequencies based on 100,000 replications.\par
$\bullet$ For Placebo\tkk-A regressions, the placebo regressor is added to
the original regression. For Placebo\tkk-R regressions, the placebo
regressor replaces the actual treatment regressor.\par 
$\bullet$ For the Monte Carlo and placebo-regression experiments,
bootstrap tests use 999 bootstrap samples, with Webb's six-point
distribution for the wild cluster bootstrap.\par
\end{table}

\Cref{tab:prf} shows $P$ values for some of the leading methods, taken
from \Cref{tab:another}, along with rejection frequencies for two
Monte Carlo experiments (for $\rho=0.00$ and $\rho=0.25$) and both
placebo\tkk-regression experiments. Interestingly, the Placebo\tkk-A
and Placebo\tkk-R results are almost identical for every method, so,
in this case, it does not seem to matter which type of placebo
regression we use. In contrast, the two Monte Carlo rejection
frequencies sometimes differ quite a lot, as the results in
\Cref{fig:A} imply.

If a method performs well, we would expect all the rejection
frequencies to be around 0.05. In such cases, the $P$ value is
probably reliable. But if a method over-rejects, we would expect its
$P$ value to be too small, and if it under-rejects, we would expect
its $P$ value to be too large. In general, smaller $P$ values in
column~1 are associated with larger rejection frequencies in the other
four columns. However, the placebo\tkk-regression results differ
substantially from the Monte Carlo ones in some cases. In particular,
the CV$_{\tn3}$ tests for the LPM and the CV$_{\tn3{\rm L}}$ tests for
the logit model over-reject quite a bit in both placebo\tkk-regression
experiments, even though they under-reject in the Monte Carlo
experiments.

The placebo\tkk-regression experiments and the Monte Carlo experiment
for $\rho=0$ all yield rejection frequencies for HC$_1$ that are very
close to 0.05. This is also the case for the default
maximum-likelihood logit standard error, denoted ``Logit ML'' in the
table. Thus it seems probable that there is actually very little
intra-cluster correlation. If so, then the most reliable methods are
the ones that ignore clustering, both of which suggest that there is
modest evidence of a treatment effect. Most of the wild bootstrap and
jackknife-based methods lead to essentially the same conclusion. The
only one that appears to provide strong evidence of an effect is the
WCLU-S bootstrap.

The other methods that apparently yield strong evidence against the
null hypothesis are the ones that combine CV$_{\tn1}$ standard errors
with the $t(11)$ distribution for both the LPM and the logit model,
which happen to be the methods that are used most commonly.
Reassuringly, both Monte Carlo experiments and both
placebo\tkk-regression experiments suggest that these methods tend to
over-reject. In contrast, the pairs cluster bootstrap suggests that
there is quite weak evidence. Again reassuringly, all four simulation
experiments suggest that it tends to under-reject.

The simulation experiments do not incorporate any form of
heteroskedasticity, because there is no evidence of it. When I
estimate separate variances for each cluster, as in \eqref{clustvar},
none of the $\hat\sigma_g^2$ differs significantly from their average.
In regression \eqref{twovars}, which checks whether treated
observations have larger variances than controls, $\hat\eta_2$ is
positive, but only by about the modest amount to be expected from the
fact that the variance of a binary variable with expectation $p$ is
$p(1-p)$, and $p$ is larger for treated observations than for
controls.

%%%%%%%%%%%%%%%%%%%%%%%%%%%%%%%%%%%%%%%%%%%%%
\subsection{Diversity in Elite Delhi Schools}
\label{subsec:india}
%%%%%%%%%%%%%%%%%%%%%%%%%%%%%%%%%%%%%%%%%%%%%

\citet{Rao_2019} studies several aspects of diversity in elite Delhi
schools. In one of his datasets, there are 2364 observations taken
from 4 grades in each of 17 schools. The ``treatment'' is the presence
of poor students in a classroom. Because of the way a policy change
requiring elite schools to make places available for poor students was
implemented, none of the grade~4 and~5 classes have any poor students,
but 24 out of 34 grade~2 and~3 classes do. The dependent variable is
binary. It equals 1 if a student volunteered for a particular charity
to raise funds for disadvantaged children; just under 26\% of the
students did~so. There are 26 explanatory variables, including school
fixed effects and grade fixed effects. The paper only estimates an
LPM, but I also estimate a logit model.

The equation I study corresponds to the first column of Table~2 in
\citet{Rao_2019}. The standard error reported in the original table is
the CV$_{\tn1}$ standard error for clustering at the school-grade
level (68 clusters), which leads to a $P$ value that is less than
$10^{-5}$, although several alternative $P$ values are also reported.
Because treatment was not assigned randomly at the school-grade level
(only grades 2 and 3 were treated, and when one grade in a school was
treated the other was usually treated as well), it seems much more
natural to cluster at the school level (17 clusters). Nevertheless,
\citet[Table~14]{Hansen_2025}, which also studies this example, reports
results only for clustering at the school-grade level.

\begin{table}[tp]
\caption{Effects of Having Poor Classmates on Volunteering for Charity}
\label{tab:poor}
\vspace*{-12pt}
\begin{center}
\begin{tabular*}{0.98\textwidth}{@{\extracolsep{\fill}}
lccccccc}
\toprule
&&\multicolumn{3}{c}{By School}&\multicolumn{3}{c}
{By School and Grade} \\
\cmidrule{3-5} \cmidrule{6-8} 
Method &Coef. &\textrm{Std.\ err.} &\textrm{$t$-stat.} &\textrm{$P$ value}
&\textrm{Std.\ err.} &\textrm{$t$-stat.} &\textrm{$P$ value} \\
\midrule
LPM HC$_1$ &0.1304 &0.0348 &3.744 &0.00019 &0.0348 &3.744 &0.00019 \\
LPM CV$_1$ &0.1304 &0.0331 &3.941 &0.00117 &0.0258 &5.048 &0.00000 \\
LPM CV$_3$ &0.1304 &0.0359 &3.633 &0.00224 &0.0377 &3.458 &0.00095 \\
LPM CV$_{\tn3}^{\rm BH}$ + $t({\rm BH})$ &0.1304 &0.0370 &3.525 &0.00397
 &0.0380 &3.432 &0.00041 \\
LPM WCU-C &0.1304 & &3.941 &0.00111 & &5.048 &0.00021 \\
LPM WCU-C (boot s.e.) &0.1304 &0.0319 &4.083 &0.00087 &0.0255 &5.117
&0.00000 \\
LPM WCU-S &0.1304 & &3.941 &0.00119 & &5.048 &0.00027 \\
LPM WCU-S (boot s.e.) &0.1304 &0.0370 &3.524 &0.00281 &0.0380 &3.436
&0.00102 \\
LPM WCR-C &0.1304 & &3.941 &0.00162 & &5.048 &0.00011 \\
LPM WCR-S &0.1304 & &3.941 &0.00177 & &5.048 &0.00014 \\
\midrule
Logit (default) &0.7105 &0.1925 &3.692 &0.00023 &0.1925 &3.692 &0.00023 \\
Logit CV$_{\tn1}$ &0.7105 &0.1883 &3.773 &0.00166 &0.1426 &4.981
&0.00000 \\
Logit CV$_{\tn3{\rm L}}$ &0.7105 &0.2087 &3.404 &0.00363 &0.2118
  &3.355 &0.00131 \\
Logit WCLU-C &0.7105 & &3.773 &0.00316 & &4.981
 &0.00000 \\
Logit WCLU-C (boot s.e.) &0.7105 &0.1818 &3.909 &0.00125 &0.1407 &5.048
 &0.00031 \\
Logit WCLU-S &0.7105 & &3.773 &0.00379 & &4.981
 &0.00041 \\
Logit WCLU-S (boot s.e.) &0.7105 &0.2152 &3.302 &0.00450 &0.2132 &3.332 
 &0.00140 \\
Logit WCLR-C &0.7105 & &3.773 &0.00179 & &4.981 &0.00012 \\ 
Logit WCLR-S &0.7105 & &3.773 &0.00188 & &4.981 &0.00016 \\ 
\bottomrule
\end{tabular*}
\end{center}
\textbf{Notes:}\par
$\bullet$ There are 999,999 bootstrap samples, which use
Rademacher weights.\par
$\bullet$ For CV$_{\tn3}^{\rm BH}$ + $t({\rm BH})$, the CV$_{\tn3}$
standard error and the degrees of freedom for the $t$ distribution are
calculated using the procedure of \citet{Hansen-jack}.\par
$\bullet$ For the logit model, CV$_{\tn3{\rm L}}$ is a
cluster-jackknife standard error based on the linearization of
\citet*{MNW-logit}.\par
$\bullet$ Unrestricted bootstrap methods marked ``boot s.e.'' use
bootstrap standard errors. The others report $P$ values for
CV$_{\tn1}$ $t$-statistics.
\end{table}

As was the case in the previous example, the effective number of
clusters is substantially smaller than the actual number. Because of
the school fixed effects, only $G^*(0)$ can be computed for clustering
at the school level; it is 11.121. For clustering by grade and school,
$G^*(0) = 28.349$ and $G^*(1) = 18.525$. These numbers suggest that
cluster-robust inference may be problematic.

\Cref{tab:poor} reports a large number of standard errors,
$t$-statistics, and $P$ values. In columns~2 through~4, clustering is
at the school level, and in columns~5 through~7 it is at the
school-grade level. In the former case, the $t$-statistics vary
modestly, from 3.524 to 4.083 for the LPM and from 3.302 to 3.909 for
the logit model. In the latter case, they vary much more, from 3.432
to 5.117 for the LPM and from 3.332 to 5.048 for the logit model. In
every case, the evidence against the null hypothesis seems to be
strong, but, based on the $P$ values, it is much stronger for some
methods than for others.

It is possible to test the hypothesis that clustering should be at the
school-grade level against the alternative that it should be at the
school level by using score\tkk-variance tests; see
\Cref{subsec:level}. The test statistic is $2.2061$, which is
asymptotically distributed as standard normal and has a one\tkk-sided
$P$ value of 0.0137. The corresponding bootstrap $P$ value, based on
99,999 bootstraps, is 0.0667. Thus, there seems to be moderately
strong evidence to support clustering at the school level.

In order to see which methods can be relied upon, I performed two
placebo\tkk-regression experiments and several Monte-Carlo
experiments. For the former, the placebo regressor was generated with
replacement by randomly treating 24 out of the 32 grade~2 and~3
classes. For the latter, I used the random-effects model \eqref{eq:RE}
in the same way as for the example of \Cref{subsec:role}. For
clustering at the school-grade level, this seems reasonably plausible,
because the school fixed effects do not absorb most of the
intra-cluster correlation. For clustering at the school level, this
model seems less plausible, but it is not clear what else to use.
Because the binary dependent variable is generated from the linear
probabilities and a normal random variable, the LPM involves some
nonlinearity, so that the fixed effects do not absorb all of the
intra-cluster correlation. However, they do absorb most of it. This
means that $\rho$ has to be quite large before there is actually much
intra-cluster correlation.

\begin{table}[tp]
\caption{Simulated Rejection Frequencies for Classmates Example}
\label{tab:raoprf}
\vspace*{-12pt}
\begin{center}
\begin{tabular*}{1.00\textwidth}{@{\extracolsep{\fill}}
lcccccccc}
\toprule
&\multicolumn{4}{c}{By School} &\multicolumn{4}{c}
{By School and Grade} \\
\cmidrule{2-5} \cmidrule{6-9} 
Method &$\rho=0$ &$\rho=0.5$
&Placebo\tkk-A &Placebo\tkk-R &$\rho=0$ &$\rho=0.25$ &Placebo\tkk-A &Placebo\tkk-R \\
\midrule
HC$_1 + t(2339)$ &0.0483 &0.0541 &0.0149 &0.0474 &0.0510 &0.4571
&0.0148 &0.0460 \\
CV$_{\tn1} + t(G-1)$ &0.0866 &0.0839 &0.0983 &0.0853 &0.1181 &0.1260
&0.1215 &0.1177 \\
CV$_{\tn3} + t(G-1)$ &0.0465 &0.0419 &0.0475 &0.0449 &0.0229 &0.0241
&0.0175 &0.0204 \\
CV$_{\tn3}^{\rm BH}$ + $t({\rm BH})$ &0.0451 &0.0406 &0.0482 &0.0426
&0.0494 &0.0499 &0.0411 &0.0427 \\
WCU-S &0.0513 &0.0466 &0.0564 &0.0507 &0.0523 &0.0518 &0.0469 &0.0432 \\
WCR-S &0.0493 &0.0490 &0.0524 &0.0508 &0.0509 &0.0500 &0.0448 &0.0426 \\
\midrule
Logit (default) &0.0494 &0.0504 &0.0201 &0.0497 &0.0518 &0.4585 
&0.0199 &0.0479 \\
Logit CV$_{\tn1}$ &0.0681 &0.0867 &0.0814 &0.0610 &0.1132 &0.1241
&0.1208 &0.1097 \\
Logit CV$_{\tn3{\rm L}}$ &0.0472 &0.0500 &0.0497 &0.0412 &0.0230 &0.0246
&0.0185 &0.0183 \\
Logit WCLU-S &0.0513 &0.0569 &0.0573 &0.0465 &0.0517 &0.0507
&0.0472 &0.0408 \\
Logit WCLR-S &0.0491 &0.0481 &0.0520 &0.0508 &0.0500 &0.0474
&0.0425 &0.0425 \\
\bottomrule
\end{tabular*}
\end{center}
\textbf{Notes:}\par
$\bullet$ Monte Carlo experiments and placebo regressions use 100,000
replications.\par
$\bullet$ Bootstrap tests use 999 bootstrap samples and the Rademacher
distribution.\par
$\bullet$ The two ``Placebo\tkk-A'' results for HC$_1$ would be identical if
it were not for variation in the random numbers. The same is true for
the two ``Placebo\tkk-R'' results and for both pairs of placebo results for
Logit (default).\par
\end{table}

\Cref{tab:raoprf} shows rejection frequencies for various tests. The
first four columns are for clustering by school, and the last four are
for clustering by school and grade. In each case, there are two sets
of Monte Carlo results and two sets of placebo\tkk-regression results.
All the wild bootstrap methods reject between 4\% and 6\% of the time,
with WCR-S arguably performing the best. Hansen's procedure also works
well in all cases.

Some of the results in \Cref{tab:raoprf} are unexpected. When there is
no clustering, the Monte Carlo experiments with $\rho=0$ suggest that
the HC$_1$ and default logit standard errors both perform very well.
The Placebo\tkk-R regressions (where the placebo regressor replaces
the actual treatment) suggest the same thing, but the Placebo\tkk-A
regressions (where the placebo regressor is added) yield quite low
rejection frequencies for these methods. This suggests that the
empirical scores must display negative intra-cluster correlation for
the Placebo\tkk-A regressions. Why this should be happening is not
entirely clear. Combining classroom fixed effects with two similar
regressors that both vary at the classroom level must be responsible.

Another odd feature is that, for all four simulations, using both
CV$_{\tn3}$ and CV$_{\tn3{\rm L}}$ standard errors leads to serious
under-rejection for school-grade clustering, but not for school
clustering. Hansen's method largely avoids this under-rejection.
Strangely, the corresponding $P$ values in \Cref{tab:poor} are
actually larger for school clustering than for school-grade
clustering.

Based on the score\tkk-variance tests and the sometimes strange
results for school-grade clustering, I conclude that clustering by
school yields more reliable results than clustering by school-grade,
even though there are only 17 clusters in the former case and 68 in
the latter. The most reliable methods seem to be the WCR-S and WCLR-S
bootstraps and Hansen's method. These yield $P$ values of 0.00177,
0.00188, and 0.00397, respectively; see \Cref{tab:poor}. Thus the
evidence that having poor classmates makes students more likely to
volunteer for charity appears to be quite strong.

%%%%%%%%%%%%%%%%%%%%%%%%%%%%%%%%%
\section{Summary and Conclusions}
\label{sec:conclusion}
%%%%%%%%%%%%%%%%%%%%%%%%%%%%%%%%%

It can be challenging to obtain reliable inferences for linear
regression models with clustered disturbances. The first issue is to
determine the correct level of clustering; see \Cref{subsec:level}.
Once this has been done, we need to estimate the variance matrix of
the parameter estimates and decide what distribution with which to
compare test statistics. The most widely-used method for individual
coefficients, based on the CV$_{\tn1}$ variance matrix \eqref{eq:CV1}
and the $t(G-1)$ distribution, very often yields $P$ values that are
misleadingly small and confidence intervals that are misleadingly
narrow. It should never be trusted blindly unless the number of
clusters ($G$) is very large and there is little heterogeneity across
clusters.

Fortunately, several better methods are available. The simplest is the
CV$_{\tn3}$, or cluster jackknife, variance matrix \eqref{eq:CV3},
which is easy to calculate and usually inexpensive if the
\texttt{summclust} package for \texttt{Stata} is used. It is always
more conservative than CV$_{\tn1}$, sometimes even too conservative.
When CV$_{\tn1}$ and CV$_{\tn3}$ yield very similar standard errors,
both of them are probably reliable. However, this is unlikely to
happen unless $G$ is quite large.

For inference about a single coefficient, the CV$_{\tn3}$ standard
error may be modified and combined with a custom degrees\tkk-of-freedom
parameter following the approach of \citet{Hansen-jack,Hansen_2025}.
The resulting $P$ values and confidence intervals may yield more
reliable inferences than ones obtained by combining CV$_{\tn3}$
standard errors with the $t(G-1)$ distribution. They perform well for
both examples in \Cref{sec:applic}.

Another method that often works very well is the WCR-S bootstrap of
\citet*{MNW-bootknife}; see \Cref{subsec:wild}. In some cases, it can
work noticeably better than the classic WCR-C bootstrap of
\citet*{CGM_2008}, although it often yields very similar results.
WCR-S bootstrap $P$ values and confidence intervals can be
surprisingly inexpensive to compute. It is tempting to believe that,
when the WCR-S bootstrap and Hansen's method yield similar $P$ values
and confidence intervals, they can be relied upon. Unfortunately, this
is not always true.

As discussed in \Cref{sec:unreliable,sec:assess}, there are cases in
which no method is reliable. Inference is particularly difficult when
the number of treated (or control) clusters is very small. In such
cases, there may be no method that can safely be used. All procedures
that use the $t(G-1)$ distribution generally over-reject, as do
bootstrap methods that do not impose a null hypothesis. In contrast,
Hansen's method and restricted wild cluster bootstrap methods
generally under-reject, sometimes to an extreme extent.

To guard against incorrect inference, it is advisable first to count
the number of clusters; the larger is $G$, the better. For models
involving treatment at the cluster level, it is also important to
count the number of treated clusters $G_1$. Assuming that treated and
control clusters are roughly the same size, the closer is $G_1/G$ to
one\tkk-half, the better. It is also valuable to calculate various
diagnostics for cluster heterogeneity; see \Cref{subsec:hetero}.
Finally, it may be a good idea to check for heteroskedasticity either
at the cluster or treatment levels, since it can affect the
performance of all inferential methods.

When competing inferential procedures yield substantively different
inferences, their performance can be evaluated by using targeted Monte
Carlo experiments (\Cref{subsec:monte}) or placebo regressions
(\Cref{subsec:placebo}). Neither of these approaches can be expected
to work perfectly, because the investigator never knows the true DGP.
Nevertheless, as the examples of \Cref{sec:applic} illustrate, they
can provide valuable information. It seems reasonable to conjecture
that, for a specific empirical application, it is generally safe to
rely on procedures which perform well in both targeted Monte Carlo
experiments and placebo regressions.

\bibliography{trust2}

@Article{Hounyo-boot,
  author = {Ulrich Hounyo and Jiahao Lin},
  title = {Wild bootstrap inference with multiway clustering and
  serially correlated time effects},
  year = {2025},
  journal = {Journal of Business \& Economic Statistics},
  url = {https://doi.org/10.1080/07350015.2025.2546454}
}

@Article{Hansen_2025,
  author  = {Bruce E. Hansen},
  title   = {Standard errors for difference-in-difference regression},
  journal = {Journal of Applied Econometrics},
  year    = {2025},
  volume  = {40},
  pages   = {291--309}
}

@Techreport{CSW_2025,
  author = {Harold D. Chiang and Yuya Sasaki and Yulong Wang},
  title = {Genuinely robust inference for clustered data},
  year = {2025},
  type = {Ar{X}iv e-prints},
  number = {2308.10138v6}
}

@Article{CHS_2024,
  author = {Harold D. Chiang and Bruce E. Hansen and Yuya Sasaki},
  title = {Standard errors for two-way clustering with serially
           correlated time effects},
  journal = {Review of Economics and Statistics},
  year = {2024}
}

@article{MNW-guide,
  author = {James G. Mac\-Kinnon and Morten {\O}. Nielsen and Matthew D. Webb},
  title = {Cluster-robust inference: {A} guide to empirical practice},
  journal = {Journal of Econometrics},
  year = {2023},
  volume = {232},
  pages = {272--299}
}

@article{MNW-logit,
  author = {James G. Mac\-Kinnon and Morten {\O}. Nielsen and Matthew D. Webb},
  title = {Cluster-robust jackknife and bootstrap inference
           for logistic regression models},
  journal = {Econometric Reviews},
  year = {2025},
  url = {https://doi.org/10.1080/07474938.2025.2515161}
}

@article{BNW_2023,
  author = {Tom Boot and Gianmaria Niccodemi and Tom Wansbeek},
  title = {Unbiased estimation of the {OLS} covariance matrix when the
  errors are clustered},
  journal = {Empirical Economics},
  year = {2023},
  volume = {64},
  pages = {2511--2533}
}

@Article{JGM-fast,
  author = {James G. Mac\-Kinnon},
  title = {Fast cluster bootstrap methods for linear regression models},
  journal = {Econometrics and Statistics},
  year = {2023},
  volume = {26},
  pages = {52--71}
}

@Article{MNW-influence,
  author = {James G. Mac\-Kinnon and Morten {\O}. Nielsen and Matthew
           D. Webb},
  title = {Leverage, influence, and the jackknife in clustered
           regression models: {R}eliable inference using summclust},
  journal = {Stata Journal},
  year = {2023},
  volume = {23},
  pages = {942--982}
}

@Article{MNW-testing,
  author = {James G. Mac\-Kinnon and Morten {\O}. Nielsen and Matthew D. Webb},
  title = {Testing for the appropriate level of clustering in linear
          regression models},
  journal = {Journal of Econometrics},
  year = {2023},
  volume = {235},
  pages = {2027--2056}
}

@Article{MNW-bootknife,
  author = {James G. Mac\-Kinnon and Morten {\O}. Nielsen and Matthew D. Webb},
  title = {Fast jackknife and bootstrap methods for cluster-robust
          inference},
  journal = {Journal of Applied Econometrics},
  volume = {38},
  year = {2023},
  pages = {671--694}
}

@Article{Cai-random,
  author = {Yong Cai},
  title = {A modiﬁed randomization test for the Level of clustering},
  journal = {Journal of Business \& Economic Statistics},
  year = {2024},
  volume = {42},
  pages = {933--945}
}

@Article{MH_2006,
  author = {Miglioretti, Diana L. and Heagerty, Patrick J.},
  title = {Marginal modeling of nonnested multilevel data using
           standard software},
  journal = {American Journal of Epidemiology},
  year = {2006},
  volume = {165},
  pages = {453--463}
}

@Techreport{Davezies_2025,
  author = {Laurent Davezies and Xavier D'Haultf{\oe}uille and Yannick 
  Guyonvarch},
  title = {Analytic inference with two-way clustering},
  type = {Ar{X}iv e-prints},
  year = {2025},
  number = {2506.20749v1}
}

@BOOK{BKW_1980,
  author = {David A. Belsley and Edwin Kuh and Roy E. Welsch},
  title = {Regression Diagnostics},
  year = {1980},
  publisher = {Wiley},
  address = {New York}
}

@Article{CW_1980,
  author = {R. Dennis Cook and Sanford Weisberg},
  title = {Characterizations of an empirical influence function for
           detecting influential cases in regression},
  journal = {Technometrics},
  year = {1980},
  volume = {22},
  pages = {495--508}
}

@Article{Efron_79,
  author = {Bradley Efron},
  title = {Bootstrapping methods: Another look at the jackknife},
  journal = {Annals of Statistics},
  volume = {7},
  year = {1979},
  pages = {1--26}
}

@Article{Efron-Stein,
  author = {Bradley Efron and Charles Stein},
  title = {The jackknife estimate of variance},
  journal = {Annals of Statistics},
  volume = {9},
  year = {1981},
  pages = {586--596}
}

@ARTICLE{Thompson_2011,
  author    = {Thompson, Samuel B.},
  title     = {Simple formulas for standard errors that cluster by 
               both firm and time},
  journal   = {Journal of Financial Economics},
  year      = {2011},
  volume    = {99},
  pages     = {1--10}
}

@Article{DMN_2019,
  author = {Antoine A. Djogbenou and James G. Mac\-Kinnon and Morten
           {\O}. Nielsen},
  title = {Asymptotic theory and wild bootstrap inference
             with clustered errors},
  journal = {Journal of Econometrics},
  year = {2019},
  volume = {212},
  pages = {393--412}
}

@Article{HansenLee_2019,
  author  = {Hansen, Bruce E. and Lee, Seojeong},
  title   = {Asymptotic theory for clustered samples},
  journal = {Journal of Econometrics},
  year    = {2019},
  volume  = {210},
  pages   = {268--290}
}

@TECHREPORT{Hansen-jack,
  author = {Bruce E. Hansen},
  title = {Jackknife standard errors for clustered regression},
  year = {2025},
  type = {Working Paper},
  institution = {University of Wisconsin}
}

@Article{RMNW,
  author = {David Roodman and James G. Mac\-Kinnon and Morten {\O}. Nielsen and
            Matthew D. Webb},
  title = {Fast and wild: Bootstrap inference in {Stata} using boottest},
  journal = {Stata Journal},
  year = {2019},
  volume = {19},
  pages = {4--60}
}

@ARTICLE{CGM_2011,
  author = {Cameron, A. Colin and Gelbach, Jonah B. and Miller, Douglas L.},
  title = {Robust inference with multiway clustering},
  journal = {Journal of Business \& Economic Statistics},
  year = {2011},
  volume = {29},
  pages = {238--249}
}

@ARTICLE{Arellano_1987,
  author = {Arellano, Manuel},
  title = {Computing robust standard errors for within groups estimators},
  journal = {Oxford Bulletin of Economics and Statistics},
  year = {1987},
  volume = {49},
  pages = {431--434}
}

@Article{MW-EJ,
  author  = {James G. Mac\-Kinnon and Matthew D. Webb},
  title   = {The wild bootstrap for few (treated) clusters},
  journal = {Econometrics Journal},
  year    = {2018},
  volume  = {21},
  pages   = {114--135}
}

@BOOK{White_1984,
  author = {White, Halbert},
  title = {Asymptotic Theory for Econometricians},
  year = {1984},
  publisher = {Academic Press},
  address = {San Diego}
}

@ARTICLE{DF_2008,
  author = {Davidson, Russell and Flachaire, Emmanuel},
  title = {The wild bootstrap, tamed at last},
  journal = {Journal of Econometrics},
  year = {2008},
  volume = {146},
  pages = {162--169}
}

@ARTICLE{DM-interp,
  author = {Davidson, Russell and Mac\-Kinnon, James G.},
  title = {The interpretation of test statistics},
  journal = {Canadian Journal of Economics},
  year = {1985},
  volume = {18},
  pages = {38--57}
}

@ARTICLE{DM_1999,
  author = {Davidson, Russell and Mac\-Kinnon, James G.},
  title = {The size distortion of bootstrap tests},
  journal = {Econometric Theory},
  year = {1999},
  volume = {15},
  pages = {361--376}
}

@ARTICLE{CGM_2008,
  author = {Cameron, A. Colin and Gelbach, Jonah B. and Miller, Douglas L.},
  title = {Bootstrap-based improvements for inference with clustered errors},
  journal = {Review of Economics and Statistics},
  year = {2008},
  volume = {90},
  pages = {414--427}
}

@Article{CSS_2017,
  author = {Carter, Andrew V. and Schnepel, Kevin T. and
    Steigerwald, Douglas G.},
  title = {Asymptotic behavior of a $t$ test robust to cluster heterogeneity},
  journal = {Review of Economics and Statistics},
  year = {2017},
  volume = {99},
  pages = {698--709}
}

@ARTICLE{MW-JAE,
  author = {Mac\-Kinnon, James G. and Webb, Matthew D.},
  title = {Wild bootstrap inference for wildly different
          cluster sizes},
  journal = {Journal of Applied Econometrics},
  year = {2017},
  volume = {32},
  pages = {233--254}
}

@Article{MW-TPM,
  author = {James G. Mac\-Kinnon and Matthew D. Webb},
  title = {Pitfalls when estimating treatment effects using
          clustered data},
  journal = {The Political Methodologist},
  year = {2017},
  volume = {24},
  pages = {20--31}
}

@ARTICLE{LZ_1986,
  author = {Liang, Kung-Yee and Zeger, Scott L.},
  title = {Longitudinal data analysis using generalized linear models},
  journal = {Biometrika},
  year = {1986},
  volume = {73},
  pages = {13--22}
}

@ARTICLE{Imbens_2016,
  author = {Guido W. Imbens and Michal {Koles\'ar}},
  title = {Robust standard errors in small samples: Some practical advice},
  journal = {Review of Economics and Statistics},
  year = {2016},
  volume = {98},
  pages = {701--712}
}

@TECHREPORT{AY-exact,
  author = {Alwyn Young},
  title = {Improved, nearly exact, statistical inference with
           robust and clustered covariance matrices using effective
           degrees of freedom corrections},
  year = {2016},
  type = {Working Paper},
  institution = {London School of Economics}
}

@ARTICLE{BM_2002,
  author = {Bell, Robert M. and McCaffrey, Daniel F.},
  title = {Bias reduction in standard errors for linear regression
           with multi-stage samples},
  journal = {Survey Methodology},
  year = {2002},
  volume = {28},
  pages = {169--181}
}

@ARTICLE{White_1980,
  author = {White, Halbert},
  title = {A heteroskedasticity-consistent covariance matrix estimator
           and a direct test for heteroskedasticity},
  journal = {Econometrica},
  year = {1980},
  volume = {48},
  pages = {817--838}
}

@ARTICLE{MW_1985,
  author = {Mac\-Kinnon, James G. and Halbert White},
  title = {Some heteroskedasticity consistent covariance matrix estimators
           with improved finite sample properties},
  journal = {Journal of Econometrics},
  year = {1985},
  volume = {29},
  pages = {305--325}
}

@Article{BCH_2011,
  author = {Bester, C. Alan and Conley, Timothy G. and Hansen,
    Christian B.},
  title = {Inference with dependent data using cluster covariance
           estimators},
  journal = {Journal of Econometrics},
  year = {2011},
  volume = {165},
  pages  = {137--151}
}

@Article{Webb_six,
  author  = {Matthew D. Webb},
  journal = {Canadian Journal of Economics},
  title   = {Reworking wild bootstrap-based inference for clustered errors},
  year    = {2023},
  volume = {56},
  pages   = {839--858}
}

@Article{IM_2016,
  author  = {Rustam Ibragimov and Ulrich K. M\"uller},
  title   = {Inference with few heterogeneous clusters},
  journal = {Review of Economics and Statistics},
  year    = {2016},
  volume  = {98},
  pages   = {83--96}
}

@Article{BDM_2004,
  author = {Bertrand, Marianne and Duflo, Esther and Mullainathan, Sendhil},
  title = {How much should we trust differences-in-differences estimates?},
  journal = {Quarterly Journal of Economics},
  year = {2004},
  volume = {119},
  pages = {249--275}
}

@ARTICLE{MNW_2021,
  author = {James G. Mac\-Kinnon and Morten {\O}. Nielsen and Matthew D.
           Webb},
  title = {Wild bootstrap and asymptotic inference with
  multiway clustering},
  journal = {Journal of Business \& Economic Statistics},
  year = {2021},
  volume = 39,
  pages = {505--519}
}

@Article{CSS_2021,
  author = {Canay, Ivan A. and Santos, Andres and Shaikh, Azeem},
  title = {The wild bootstrap with a ``small'' number of ``large'' clusters},
  year = {2021},
  journal = {Review of Economics and Statistics},
  volume = {103},
  pages = {346--363}
}

@TechReport{NAAMW_2020,
  author = {Gianmaria Niccodemi and Rob Alessie and Viola Angelini
            and Jochen Mierau and Tom Wansbeek},
  title = {Refining clustered standard errors with few clusters},
  year = {2020},
  institution = {University of Groningen},
  type = {Working Paper},
  number = {2020002-EEF}
}

@ARTICLE{Tukey_1958,
  author = {John W. Tukey},
  title = {Bias and confidence in not quite large samples},
  journal = {Annals of Mathematical Statistics},
  year = {1958},
  volume = {29},
  pages = {614}
}

@TechReport{MNW-twoway,
  author = {James G. Mac\-Kinnon and Morten {\O}. Nielsen and Matthew D. Webb},
  title = {Jackknife inference with two\tkk-way clustering},
  year = {2026},
  type = {ar{X}iv e-prints},
  number = {2406.08880}
}

@ARTICLE{PS_2020,
  author = {Catherine Porter and Danila Serra},
  title = {Gender Differences in the Choice of Major: {T}he Importance of
          Female Role Models},
  journal = {American Economic Journal: Applied Economics},
  year = {2020},
  volume = {12},
  pages = {226--254}
}

@ARTICLE{Rao_2019,
  author = {Gautam Rao},
  title = {Familiarity Does Not Breed Contempt: 
          Generosity, Discrimination, and Diversity in {D}elhi Schools},
  journal = {American Economic Review},
  year = {2019},
  volume = {109},
  pages = {774--809}
}

@BOOK{MHE_2008,
  author = {Joshua D. Angrist and Jorn-Steffen Pischke},
  title = {Mostly Harmless Econometrics: An Empiricist's Companion},
  publisher = {Princeton University Press},
  year = {2008}
}

@ARTICLE{CT-persist,
  author = {Timothy G. Conley and Morgan Kelly},
  title = {The standard errors of persistence},
  journal = {Journal of International Economics},
  year = {2025},
  volume = {153},
  pages = {Article 104027}
}

@ARTICLE{Vogel_2024,
  author = {Chen, Kaicheng and Vogelsang, Timothy J.},
  title = {Fixed-$b$ asymptotics for panel models with two-way
          clustering},
  journal = {Journal of Econometrics},
  year = {2024},
  volume = {244},
  number = {105831},
  pages = {1-25}
}
\addcontentsline{toc}{section}{\refname}

\end{document}